\documentclass[twocolumn]{aastex631}
\usepackage{subfigure}

\begin{document}

\title[Optical variability of {\it Swift}/ BAT AGN]{Exploring the Origins of Optical Variability in AGNs: Correlations with Black Hole Properties, X-ray, and Radio Emission}

\author[0000-0002-3277-6335]{Vivek Kumar Jha}
\affiliation{National Centre for Radio Astrophysics, Tata Institute of Fundamental Research, Post Bag 3, Ganeshkhind, Pune, 411007; India}
\affiliation{Manipal Centre for Natural Sciences, Manipal Academy of Higher Education (MAHE), Manipal, Karnataka,  576104; India}

\author[0000-0002-5123-972X]{Debbijoy Bhattacharya}
\affiliation{Manipal Centre for Natural Sciences, Manipal Academy of Higher Education (MAHE), Manipal, Karnataka,  576104; India}

\author[0000-0002-3163-4941]{Hum Chand}
\affiliation{Department of Physics and Astronomical Sciences, Central University of Himachal Pradesh, Dharamshala, 176215; India}

\correspondingauthor{Vivek Kumar Jha}
\email{vivekjha.aries@gmail.com}

\begin{abstract}

We study the optical variability characteristics of Active Galactic Nuclei (AGN) from the  Swift Burst Alert Telescope (BAT) AGN catalogue by utilising approximately five years of optical light curves from the Zwicky Transient Facility (ZTF) survey. We investigate dependencies of the long-term optical variability amplitudes and timescales on (i) supermassive black hole (SMBH) mass, luminosity, and Eddington ratio to explore the influence of accretion disk dynamics and radiative processes; (ii) X-ray properties, such as spectral photon indices and fluxes, to study the effect of high-energy emission mechanisms; and (iii) radio characteristics, such as integrated fluxes and radio loudness, which indicate jet activity. Our findings confirm a positive correlation between the variability time scale and both the SMBH mass and luminosity, suggesting that these physical parameters significantly impact the optical variability timescale. Conversely, no significant dependence is found between optical variability and X-ray properties, indicating that high-energy processes may not substantially influence long-term optical variability. Additionally, a weak anti-correlation between optical variability and radio parameters suggests that jet activity has a negligible effect on causing long-term AGN variability. These results support the hypothesis that long-term optical variability in AGN is primarily governed by thermal emission from the accretion disk. Further investigations with larger samples are essential to refine these correlations and develop robust physical models integrating black hole properties, accretion disk physics, and multi-wavelength radiative transfer.

\end{abstract}




\section{Introduction} \label{sec:intro}

Active Galactic Nuclei (AGN) are among the most variable sources in the Universe, exhibiting changes in their luminosity and emission spectra across the electromagnetic spectrum \citep{variability_review_ulrich,2002MNRAS.329...76H,2004ApJ...601..692V}. These changes span a wide range of timescales, from minutes to years, reflecting the various physical processes occurring in the vicinity of supermassive black holes. Several different mechanisms have been proposed to explain AGN variability, including change of the accretion rate onto the black hole, variations in the structure of the accretion disk, and the presence of clouds or other structures in the vicinity of the black hole that can occult the central source \citep[e.g., see][etc.]{2008MNRAS.387L..41L,2018ApJ...866...74S,2020ApJ...891..178S}. Variability can also be used to probe the accretion disk's inner regions, as these objects are too small to be resolved spatially.

Variability in any wavelength offers critical insights into the dynamic processes governing the time evolution of emission in AGN. Such variability serves as a diagnostic tool to understand the characteristic timescales and physical mechanisms that drive fluctuations in different emission regions. Connecting the AGN variability with the physical parameters can help understand the behaviour of the ensemble population of AGN \citep[e.g., see][]{2010ApJ...716L..31A,2023NatAs...7..473T}, while sometimes, the variability can be used to distinguish AGN from other astronomical objects in extensive surveys \citep{2011AJ....141...93B}. Different approaches are used to study the variability trends: the structure-function (SF) \citep{2016_kozlowski_sf} and Power Spectrum Density (PSD) modelling of the light curves, which work in frequency domain \citep{Simm2016} while the excess variance \citep{vaughan2003} in the light curves can be used to study the variability in the time domain. Recently, the AGN stochastic fluctuation has been modeled as a Damped Random Walk (DRW), the simplest approximation of the Continuous-time AutoRegressive Moving Average (CARMA) process \citep{Kelly2009, MacLeod2010,zu2013}. The characteristic time scale obtained through the DRW modelling ($\tau$)and its correlation with the SMBH mass suggests that the SMBH mass could influence the variability of the AGN \citep{2021Sci...373..789B}.

With the advent of all-sky surveys, studying the properties of multiple AGNs simultaneously has become possible. ZTF is one such all-sky survey that provides continuous observations in the g,r, and i bands spanning over 5 years \citep{Bellm2019}. ZTF light curves have been used in AGN studies in recent works, for example, measurement of the accretion disk sizes \citep{guo2022, Jha2022} and deep learning based modeling of the quasar variability phenomenon\citep{2020ApJ...903...54T}. \citet{2022ApJ...936..132Y} explored the modelling of AGN light curves beyond the DRW model, and found that the damped harmonic oscillator (DHO) process can provide more insights into AGN variability. \citet{2023Ap&SS.368...68W} found that there is a luminosity and SMBH mass dependence on the variability amplitudes. \citet{2023ApJ...951...93C} found a correlation between the SMBH mass and $\tau$ based on the infrared (IR) variability. Variability and its correlation with the spectroscopic parameters, such as emission line widths and shapes, have also been observed \citep{2021ApJ...911..148K}. Despite the recent works, the dependence of the variability amplitude on the luminosity, accretion rate, and emission line strength is unclear and inconsistent among different studies \citep[e.g., see][etc.]{2018ApJ...864...87S,2019ApJS..242...10S, Lu2019, 2020MNRAS.495.1403X,2021ApJ...907...96S, 2022A&A...664A.117D,2022MNRAS.514..164S,2023MNRAS.526.6078A,2024A&A...684A.133A}

\begin{figure}

    \includegraphics[width=9cm,height=7cm]{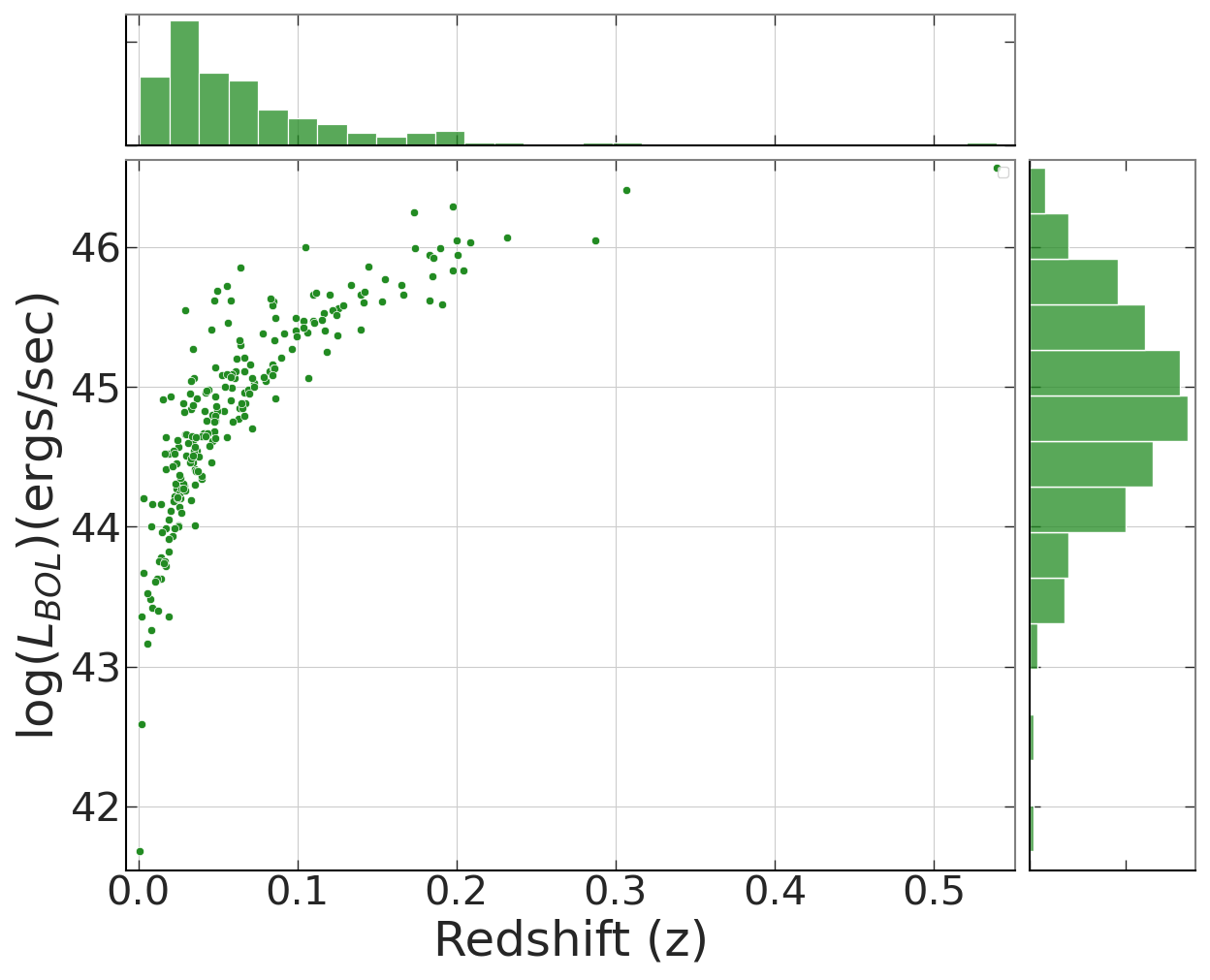}
    \caption{The distribution of the bolometric luminosity and redshift for the sample of the {\it Swift/}BAT AGN sample studied here.}
    \label{fig:lz_plot}
\end{figure}

Connecting the variability with parameters obtained from multiple wavelengths can shed essential insights into its physical origin. For example, the detection of radio emission can provide valuable insights into the presence and properties of these collimated outflows of matter and energy ejected from the vicinity of the supermassive black hole \citep{2019NatAs...3..387P}. Radio-loud AGN, characterised by the presence of powerful jets, may be viewed along an axis aligned with the jet direction, while the dusty torus in our line of sight may obscure radio-quiet AGN. Studies have looked for the correlation between the radio and optical emission of the light curves and found that radio emission can drive the UV optical variability \citep[e.g. see][etc.]{1995AJ....110..529C,2022MNRAS.512..296L} It remains to be seen how the long term UV/Optical variability is connected to the radio flux and loudness.

X-ray emission from AGN originates from the hot accretion disk and/or corona surrounding the supermassive black hole \citep{1991ApJ...380L..51H,2016ApJ...821L...1N}, while optical emission originates from the cooler outer regions of the accretion disk. However, obscuration by dust and gas along the line of sight can significantly bias the interpretation of X-ray spectral properties, such as the photon index ($\Gamma$), and limit our understanding of the intrinsic variability of the central engine \citep{2014MNRAS.437.3550M}. A correlation in variability across the optical and X-ray bands could imply coupling between disk and corona dynamics, perhaps through the reprocessing of high-energy X-ray photons by the accretion disk or through a common variability-driving mechanism, such as changes in the accretion rate or magnetic processes affecting both emission regions. Understanding the nature of these connections is essential for refining AGN variability models and probing the physics of multi-wavelength emission processes in AGNs.

The Swift Burst Alert Telescope (BAT)-AGN catalogue provides a unique sample of hard X-ray selected AGN, to study their properties in conjunction with other multi-wavelength parameters \citep{2009ApJ...690.1322W}. Recent studies have probed the long-term X-ray variability of active galactic nuclei (AGNs) using the BAT AGN catalogue \citep{soldi2024,2020ApJ...896..122L}. The 105-month Swift-BAT survey has identified 1632 hard X-ray sources, including a substantial number of Seyfert AGN, which are crucial for understanding the X-ray emission mechanisms. \citet{2017MNRAS.470..800T} studied the connection between X-ray spectral index, $\Gamma$ and accretion rate and found a weak correlation between the two parameters. \citet{2022ApJ...939L..13A} analyzed the Eddington ratio distribution and its dependence on neutral Hydrogen column density, showing that obscured AGN dominate at low Eddington ratios, while unobscured AGN dominate at high Eddington ratios. \citet{2023MNRAS.526.1687T} studied the X-ray variability and found a correlation of the X-ray excess variance with SMBH mass. \citet{2024A&A...685A..50P} using 100 brightest AGN from this sample, studied the PSD of X-ray light curves and found that the power spectrum of Seyfert 1 and Seyfert 2 are similar and that the PSD amplitude decreases with increase in accretion rate. However, the link between long-term X-rays and optical variability remains to be determined. Even though the short-time flicker in X-ray light curves can arise in the innermost regions, the long-term variability in X-ray should be affected by the propagating optical/UV fluctuations inwards to the disk \citep{1997MNRAS.292..679L}.

 \begin{figure*}
\centering

\subfigure{\includegraphics[width=8.5cm,height=5cm]{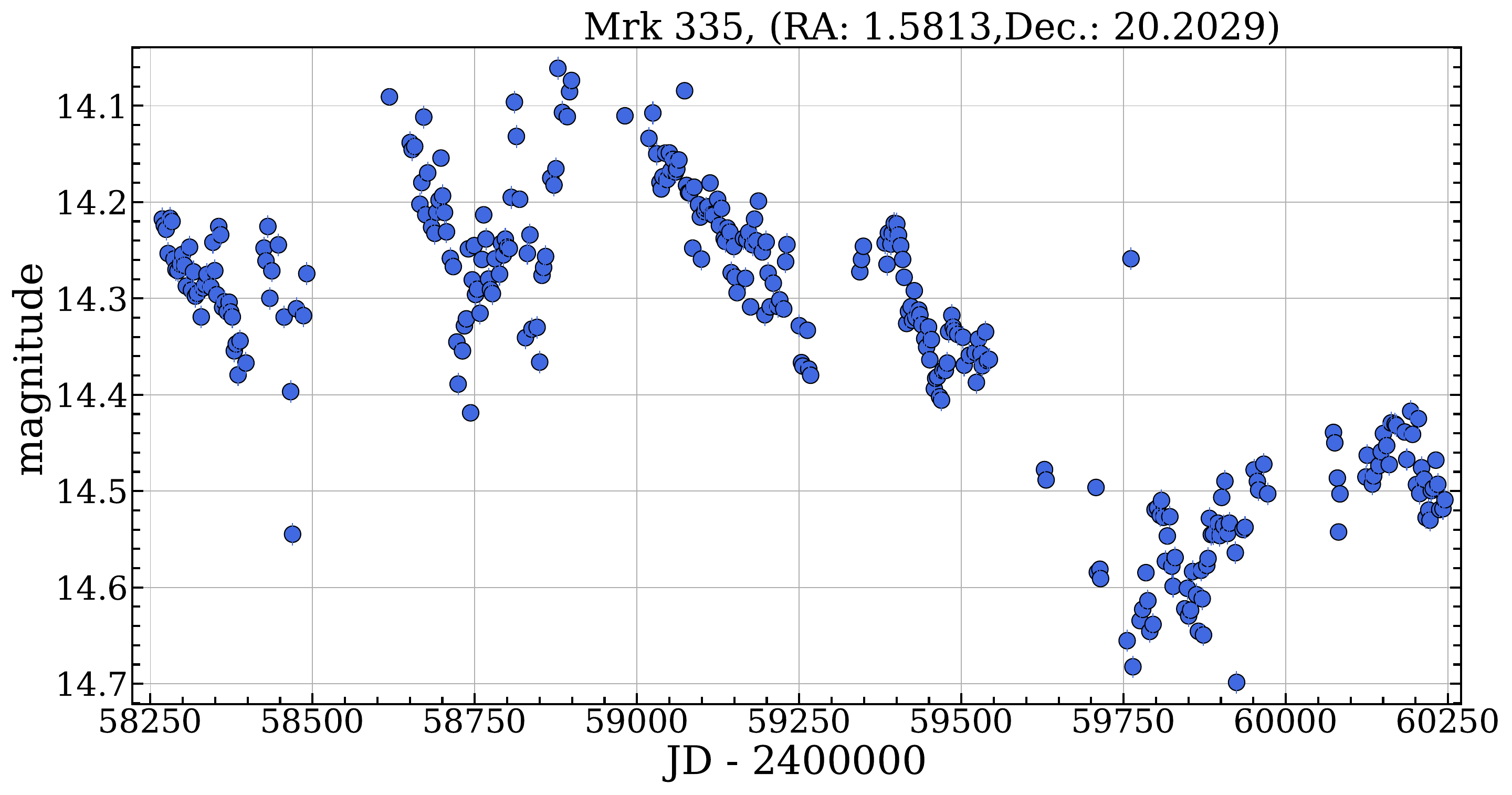}}
\subfigure{\includegraphics[width=8.5cm,height=5cm]{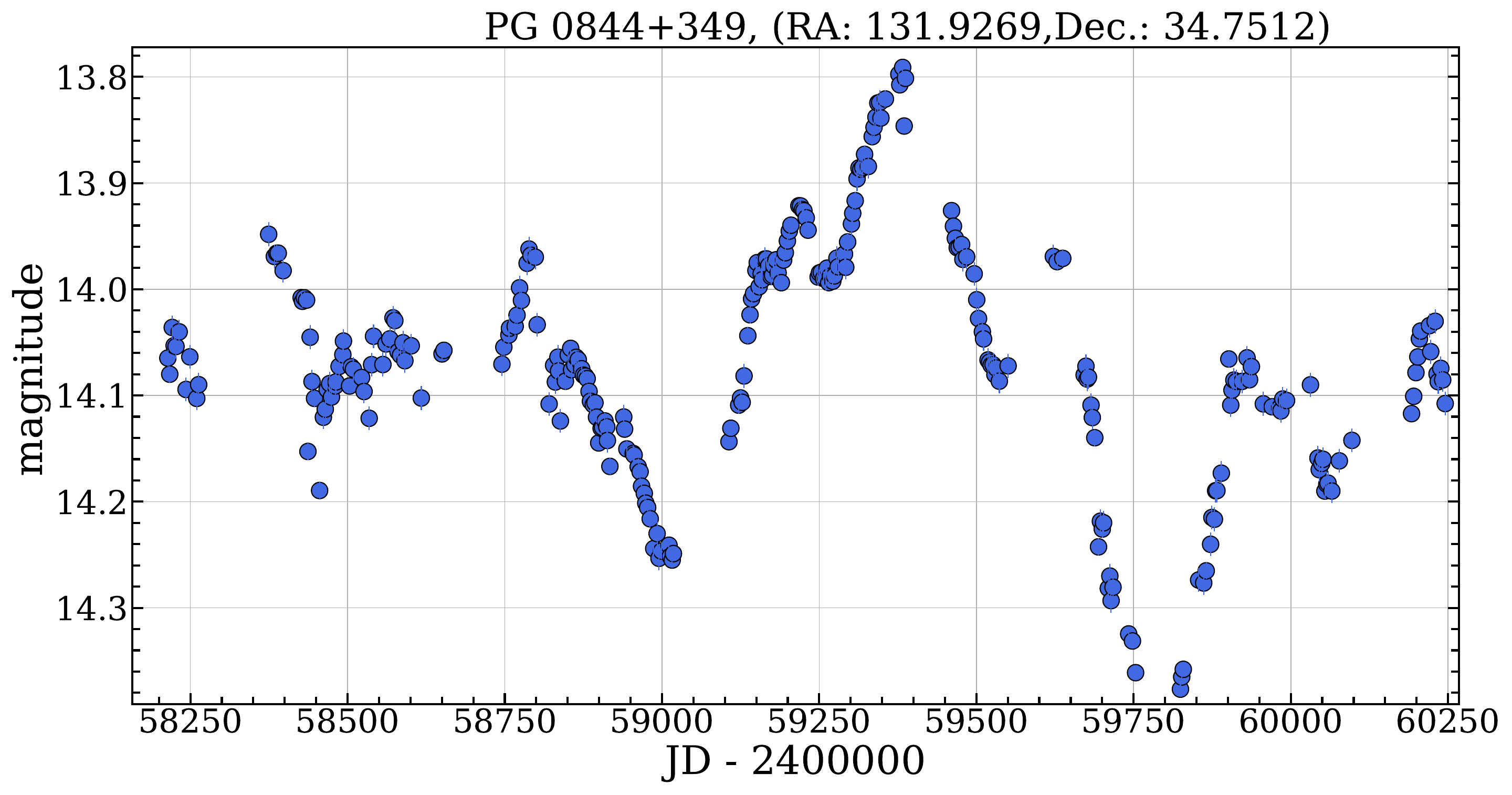}}
\subfigure{\includegraphics[width=8.5cm,height=5cm]{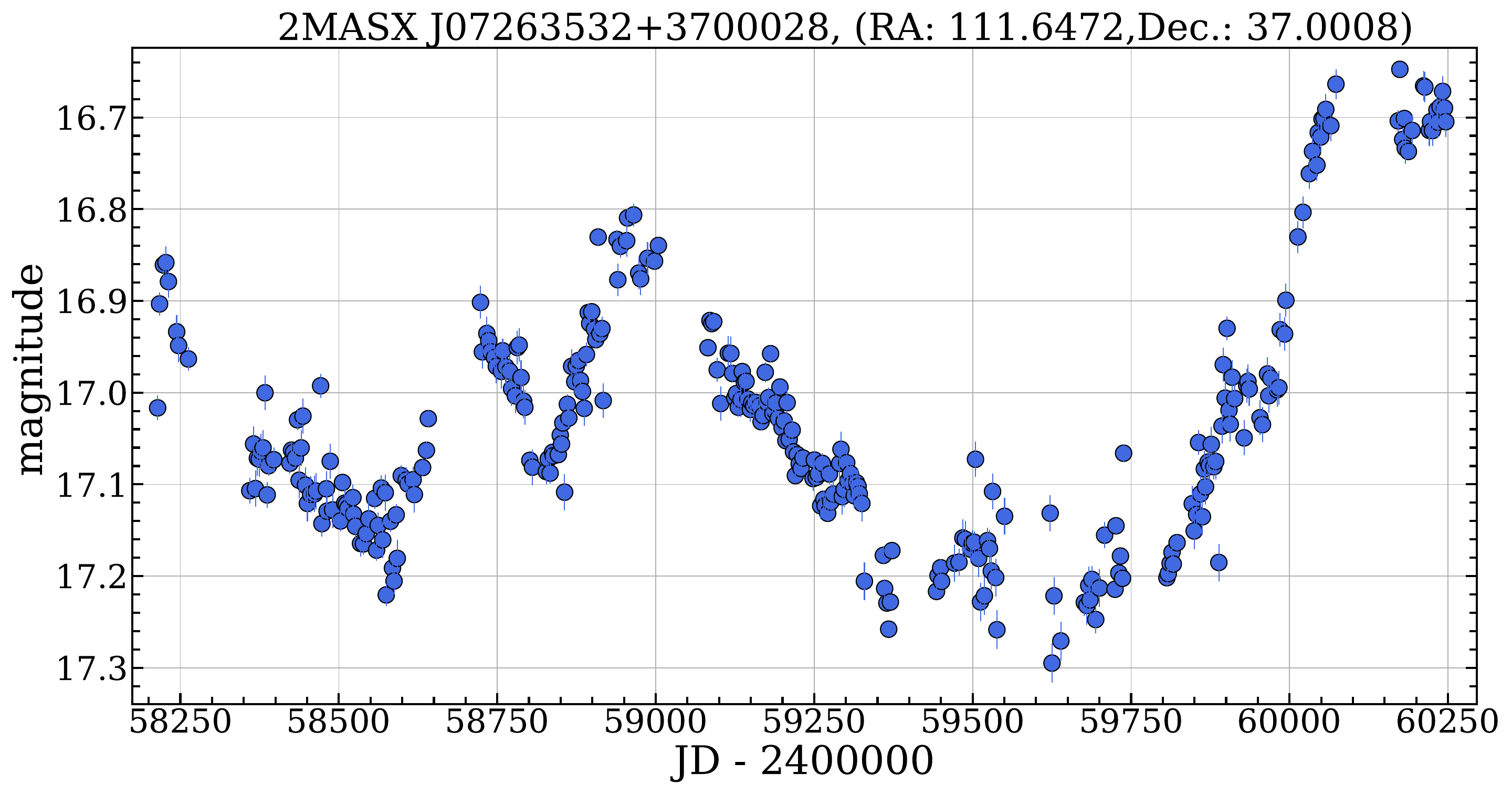}}
\subfigure{\includegraphics[width=8.5cm,height=5cm]{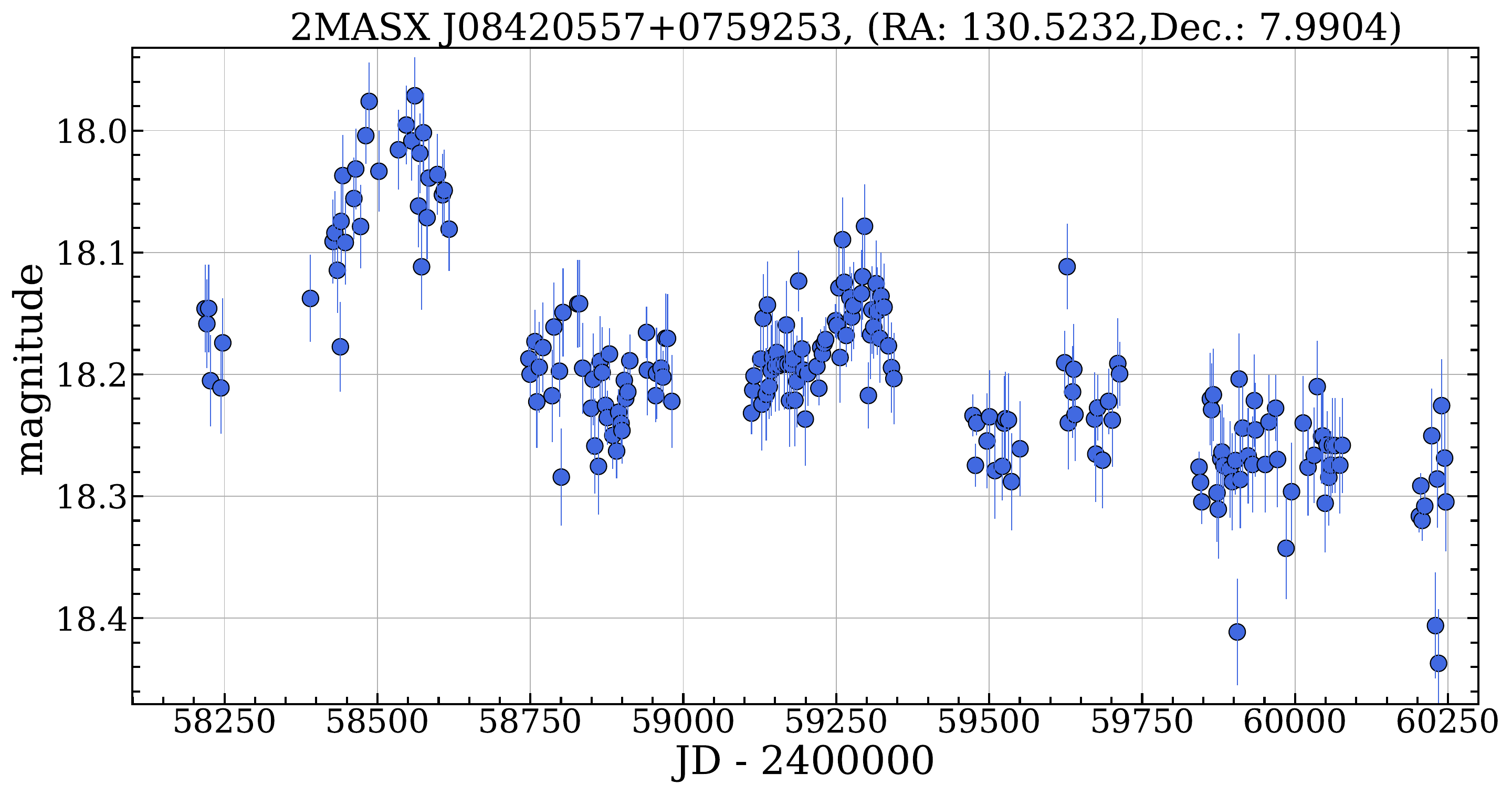}}
\subfigure{\includegraphics[width=8.5cm,height=5cm]{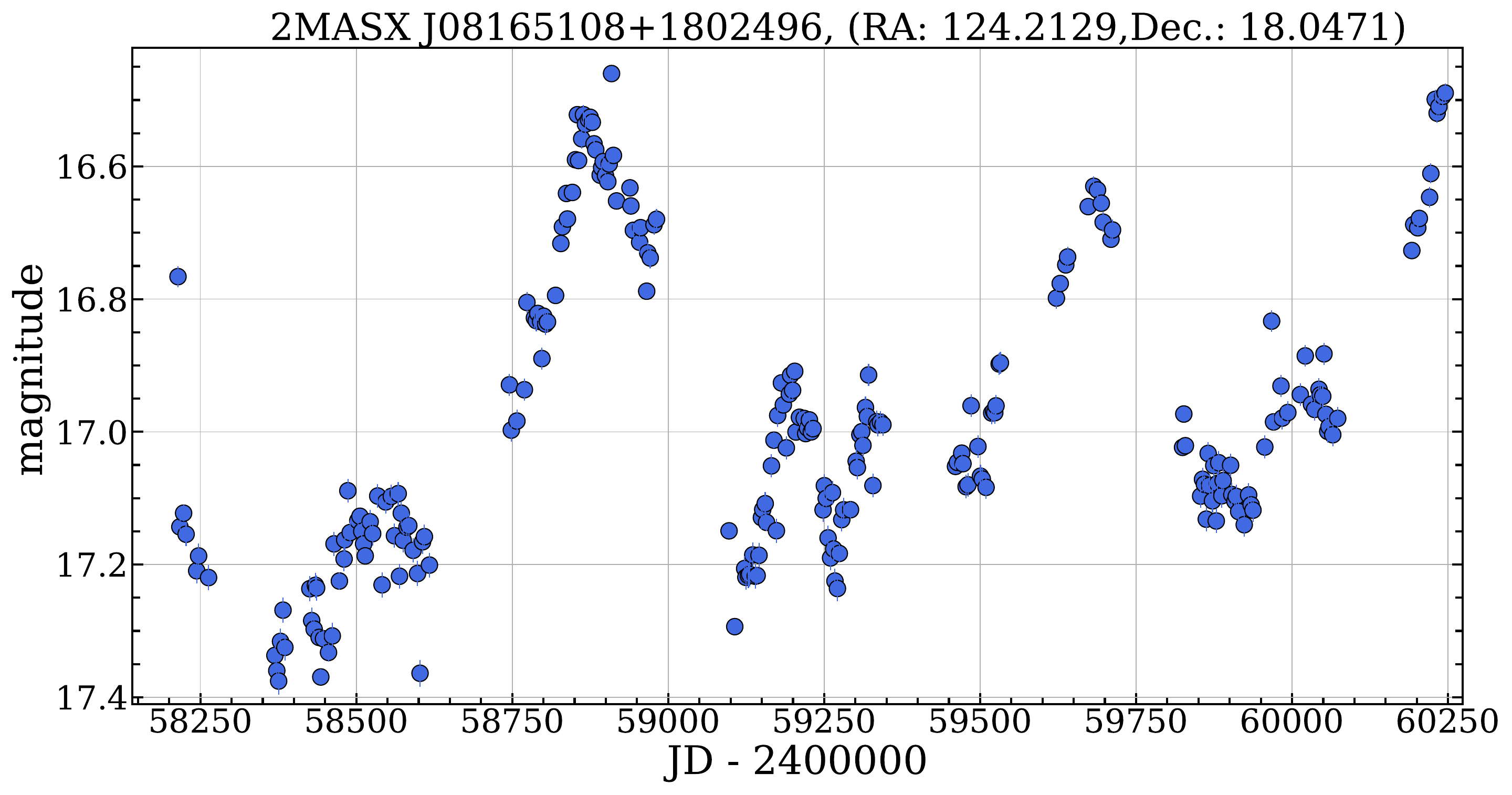}}
\subfigure{\includegraphics[width=8.5cm,height=5cm]{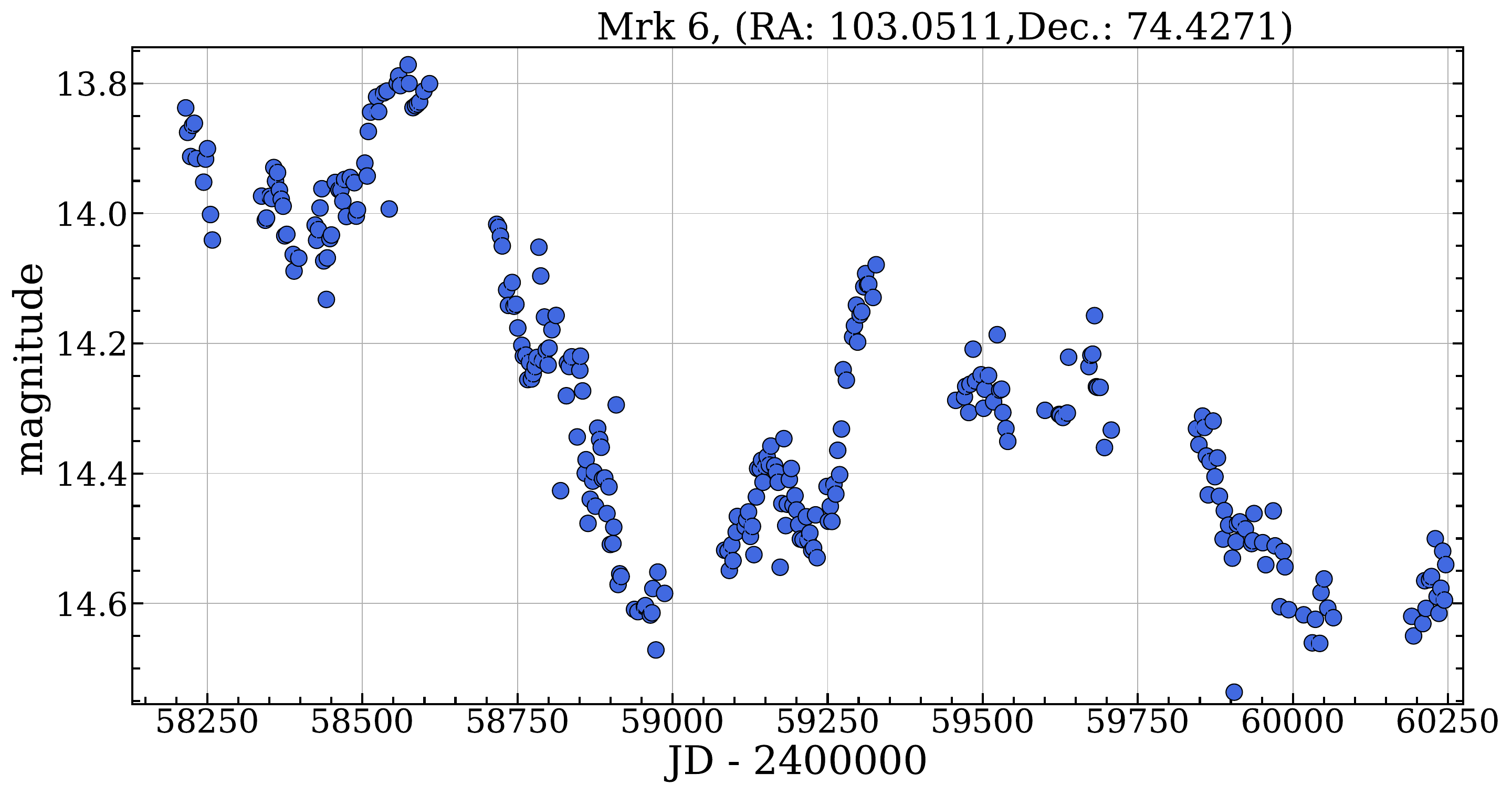}}

\caption{The ZTF r-band light curves for 6 of the sources studied in this work. The common name for the source, along with right ascension (RA) and Declination (Dec.), are noted on the top of each panel.}
\label{light_curves}
\end{figure*}

In this work, we study the correlation of the optical variability in the light curves from the ZTF survey with the multi-wavelength parameters spanning optical, X-ray and radio wavelengths. Here we aim to unveil crucial aspects of AGN variability and its dependence on the connection between the accretion disk, the supermassive black hole, and the surrounding environment. This can also shed light on the unification scheme, providing new insights into the diverse nature of AGN. The organisation of this paper is as follows: In Section \ref{section2}, we describe the sample selection and the data acquisition for this study. In Section \ref{section3}, we outline the methods to quantify the variability and the relation with the physical parameters. In Section \ref{section4}, we report the results from this analysis. In Section \ref{section5}, we discuss the implications of this work, followed by a summary in Section \ref{section6}.

 \begin{figure*}

\hspace{-1cm}
\subfigure{\includegraphics[width=6cm,height=5.5cm]{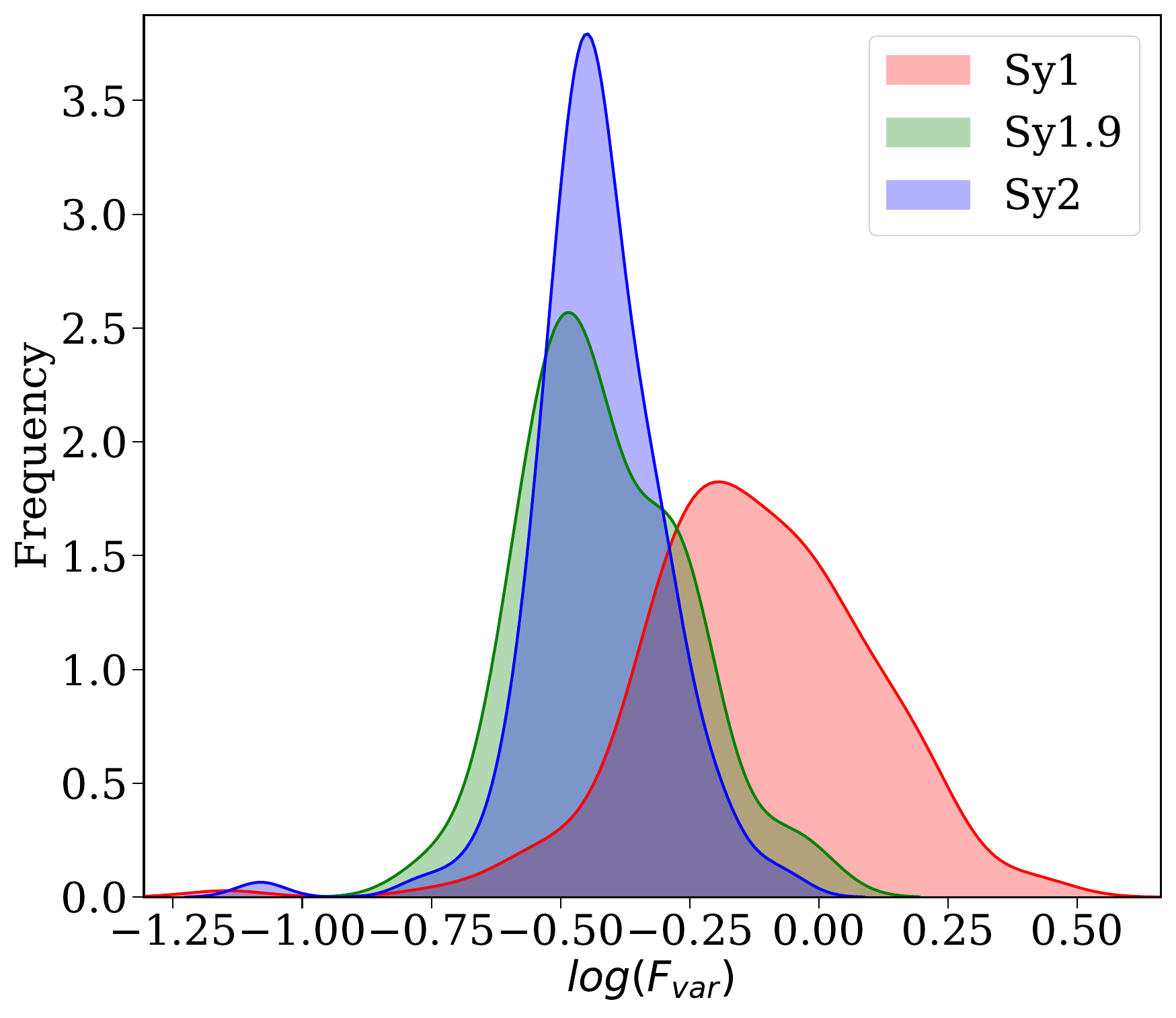}}
\subfigure{\includegraphics[width=6cm,height=5.5cm]{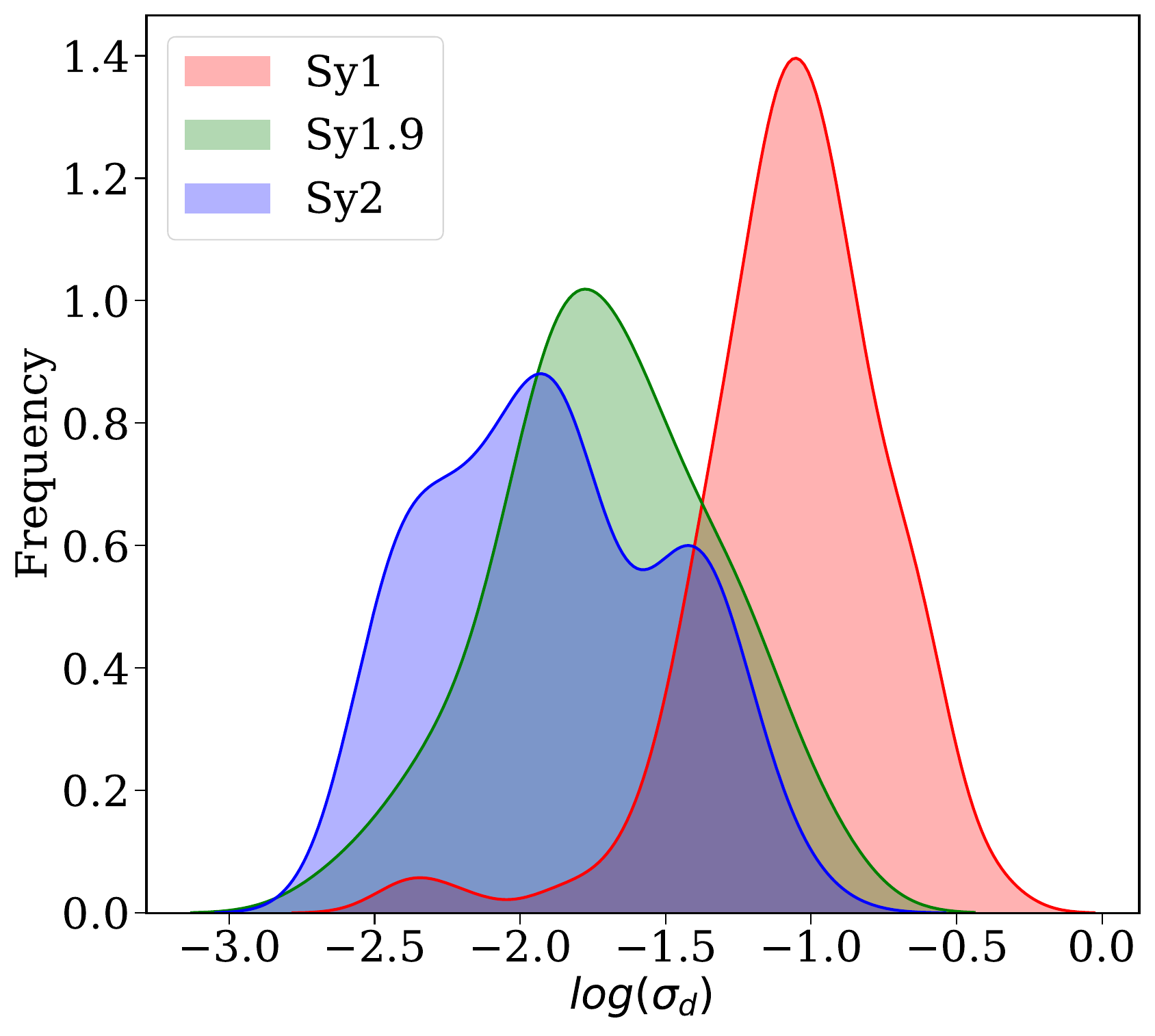}}
\subfigure{\includegraphics[width=6cm,height=5.5cm]{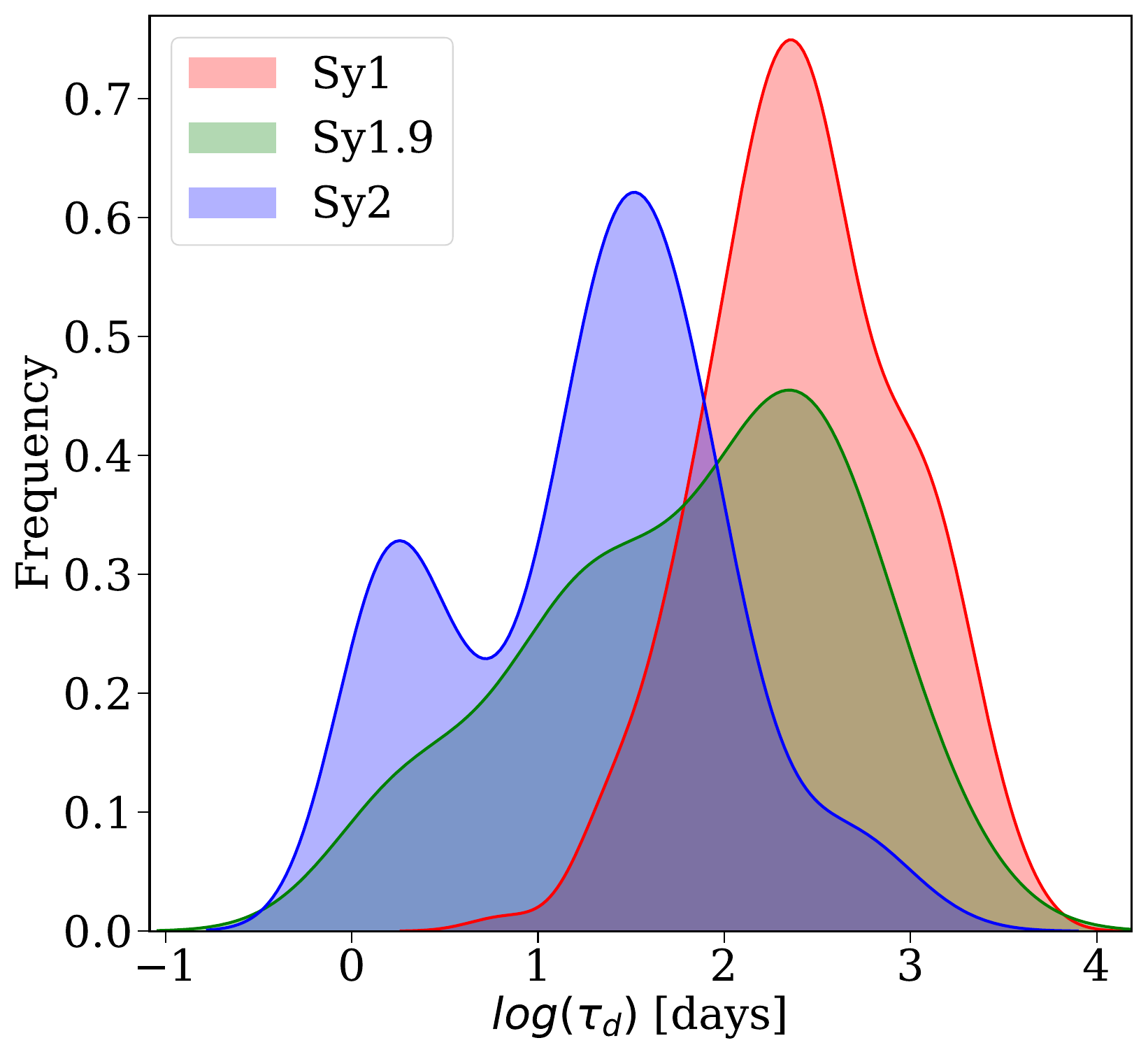}}

\caption{Distribution of the logarithm of excess variance $F_{var}$ (left), the variability amplitudes $\sigma$ (center) and the damping timescale $\tau$ (right) for our sample. The sources are divided into three types (Sy1, Sy1.9 and Sy2) based on the classification available in \citet{koss2022} for the sample of AGN from the BAT 70-month catalogue. For clarity, we merged the Sy1.2, 1.5, and 1.8 sources into the Sy1 category for further analysis.}
\label{hist_params}
\end{figure*}

\begin{table}
    \centering
    \caption{Details of AGN from SWIFT BAT 105-month Catalogue and ZTF Search Results}
    \begin{tabular}{lcc}
        \toprule
        \textbf{AGN Type} & \textbf{SWIFT BAT Sources} & \textbf{ZTF Sources} \\
        \hline
        Seyfert 1.0 (Sy1)  & 160 & 110 \\
        Seyfert 1.2 (Sy1.2) & 100 & 71  \\
        Seyfert 1.5 (Sy1.5) & 110 & 79  \\
        Seyfert 1.8 (Sy1.8) & 10  & 8   \\
        Seyfert 1.9 (Sy1.9) & 138 & 70  \\
        Seyfert 2 (Sy2)   & 310 & 190 \\
      \hline
        \textbf{Total} & \textbf{828} & \textbf{528} \\
       \hline
    \end{tabular}
\label{search_results}
\end{table}

\section{The sample and data}
\label{section2}

We obtained our sample of AGN from the Swift-BAT 105-month catalogue, which includes data from 105 months of observations by the Burst Alert Telescope (BAT) onboard the Swift observatory. This catalogue comprises 1632 hard X-ray sources detected in the 14-195 keV band, with 828 identified as Seyfert galaxies. Out of these, 310  are classified as Type 2 AGN. The remaining 518 sources fall into the Seyfert 1  category: 160 as Seyfert 1.0, 100 as Seyfert 1.2, 110 as Seyfert 1.5, 10 as Seyfert 1.8, and 138 as Seyfert 1.9. The catalogue provides detailed information on each source, including unique identification numbers and names, Right Ascension and Declination coordinates in the J2000.0 epoch, and Galactic longitude and latitude. Additionally, it offers photometric data such as flux measurements, along with associated errors in the 14-195 keV band and spectral photon indices \citep{koss2022}.

To get the optical light curves, we utilised data from the Zwicky Transient Facility (ZTF), an optical time-domain survey designed to monitor the entire northern sky \citep{Bellm2019}. ZTF uses a 1.2-meter Samuel Oschin Telescope at the Palomar Observatory and is equipped with a 47-square-degree field-of-view camera, enabling it to scan the sky rapidly and repeatedly. The survey captures images in three bands: g, r, and i. We search the sources in the ZTF data release (DR) 20 by using a search radius of 5 arcseconds on the counterpart right ascension (RA) and declination (Dec) positions. We searched for light curves in all three bands available in the ZTF $-$ for the g, r, and i bands. The ZTF light curves are significantly shorter in the i band, while the length of these light curves is comparable in the g and r bands. Our search yielded 110 Seyfert 1, 71 Seyfert 1.2, 79 Seyfert 1.5, 8 Seyfert 1.8, 70 Seyfert 1.9, and 190 Seyfert 2 galaxies in the r band (see Table \ref{search_results}). We applied an additional criterion, requiring each source to have at least 100 data points. We applied a 3-sigma clipping on all the light curves to filter out the outlier points. We also removed repeated observations with multiple object identifiers (OIDs) at each location. In case of multiple OIDs present at a single location, we selected the observation with the maximum number of data points. We obtained r-band ZTF optical light curves for 528 sources in our sample. Figure \ref{fig:lz_plot} presents our sample in the luminosity redshift plane, and Figure \ref{light_curves} presents ZTF r band light curves for 6 sources in our sample.

\begin{figure}

    \includegraphics[width=9cm,height=8.5cm]{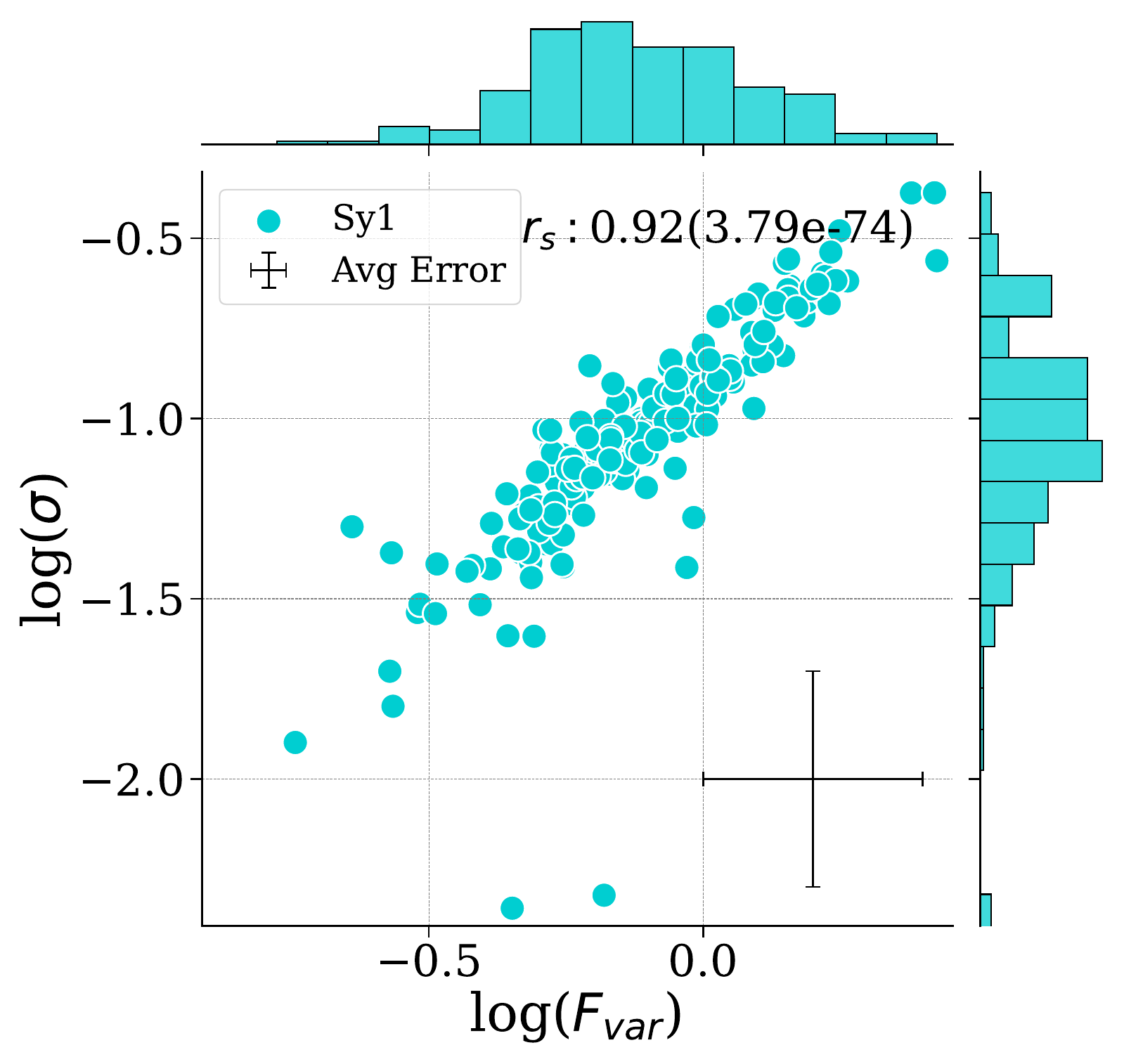}
    \caption{The relation between the excess variance ($F_{var}$) and the DRW variability amplitude ($\sigma$) for the 227 Type-1 AGN from our sample. The Spearman rank correlation coefficient is shown on the top of the panel, with the $p_{null}$ value in the bracket. The median errorbar on both the parameters is shown in bottom right.}
    \label{comparison_fvar}
\end{figure}

 \begin{figure*}

\hspace{-1cm}
\subfigure{\includegraphics[width=6cm,height=5.5cm]{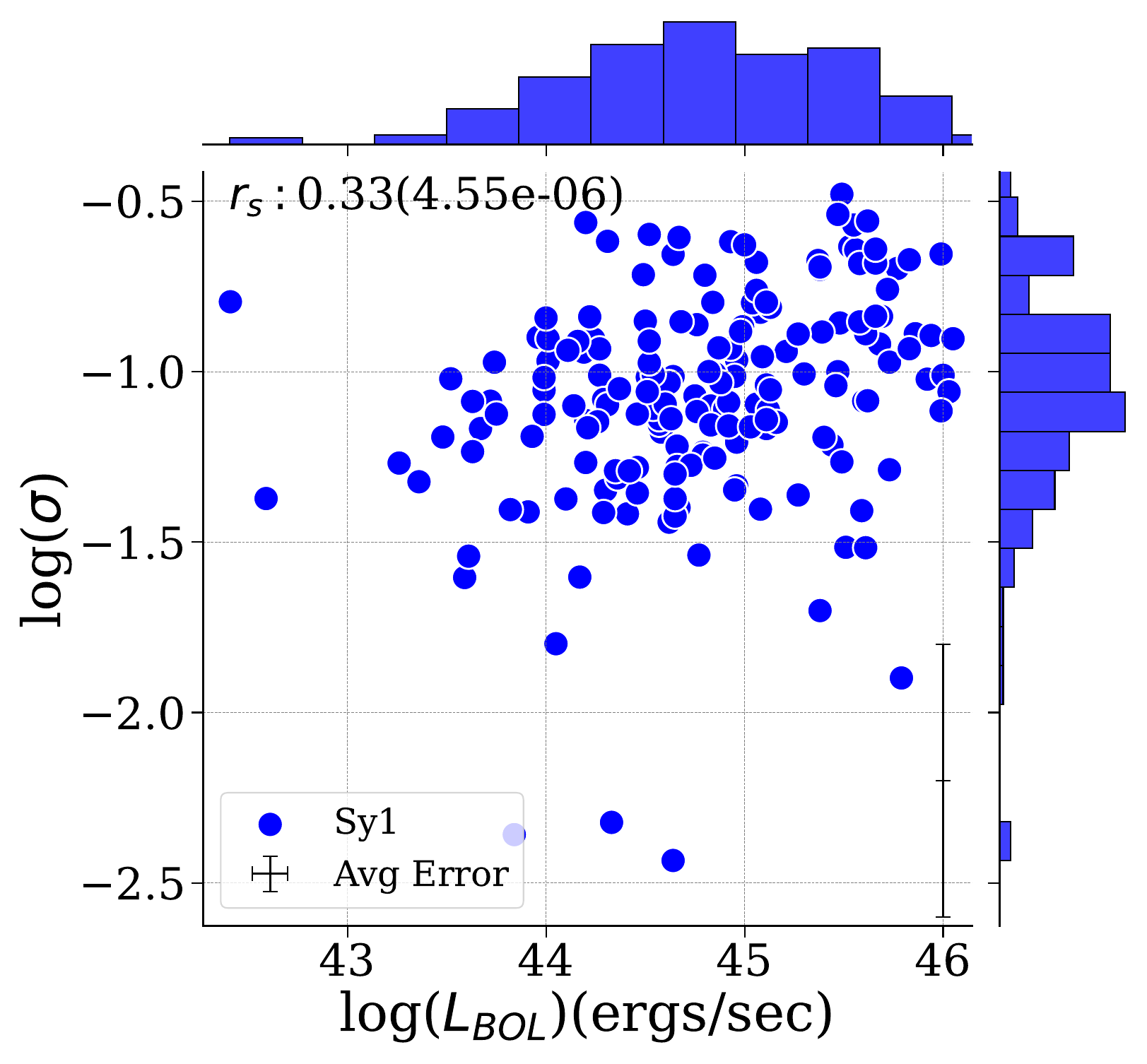}}
\subfigure{\includegraphics[width=6cm,height=5.5cm]{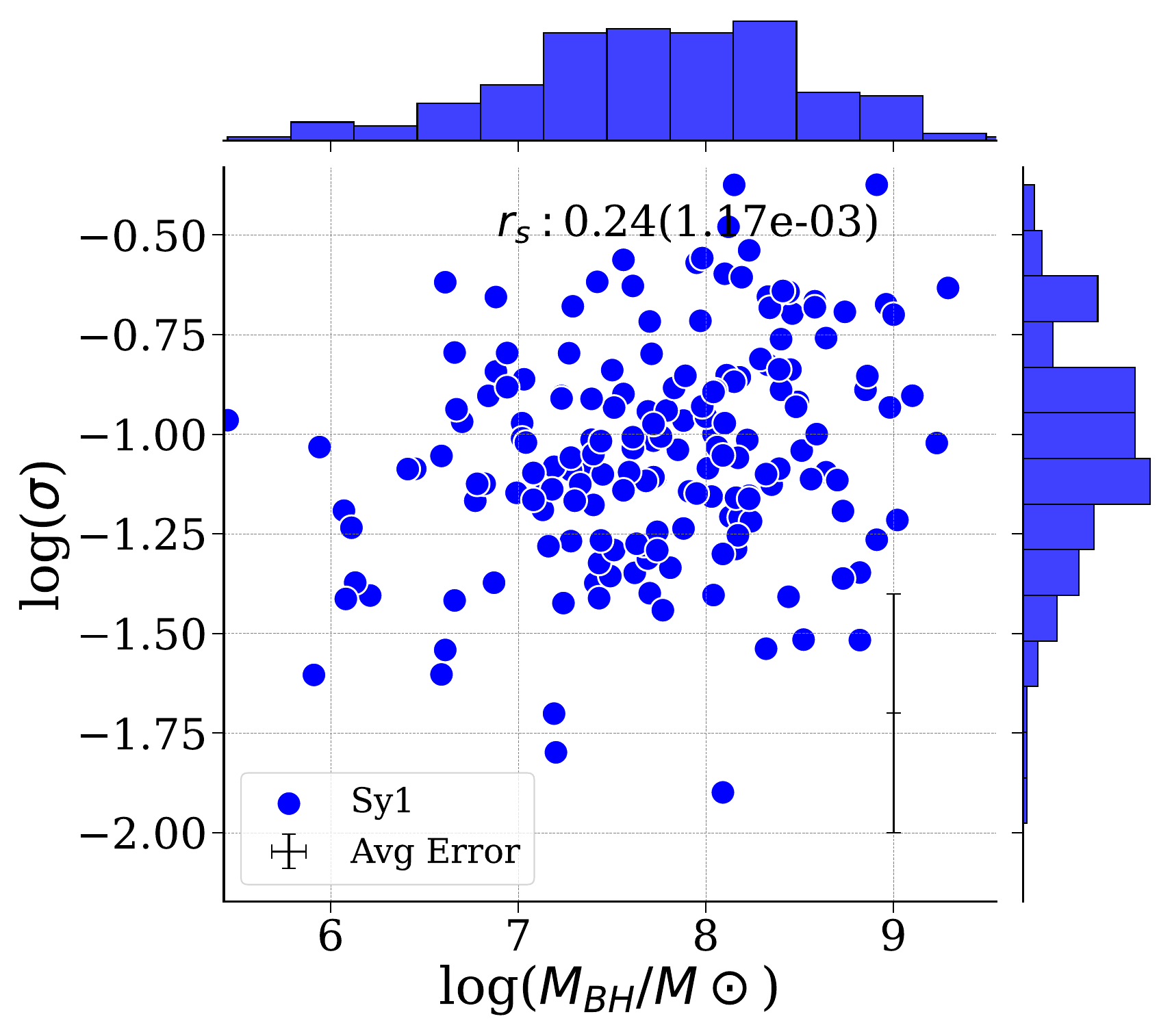}}
\subfigure{\includegraphics[width=6cm,height=5.5cm]{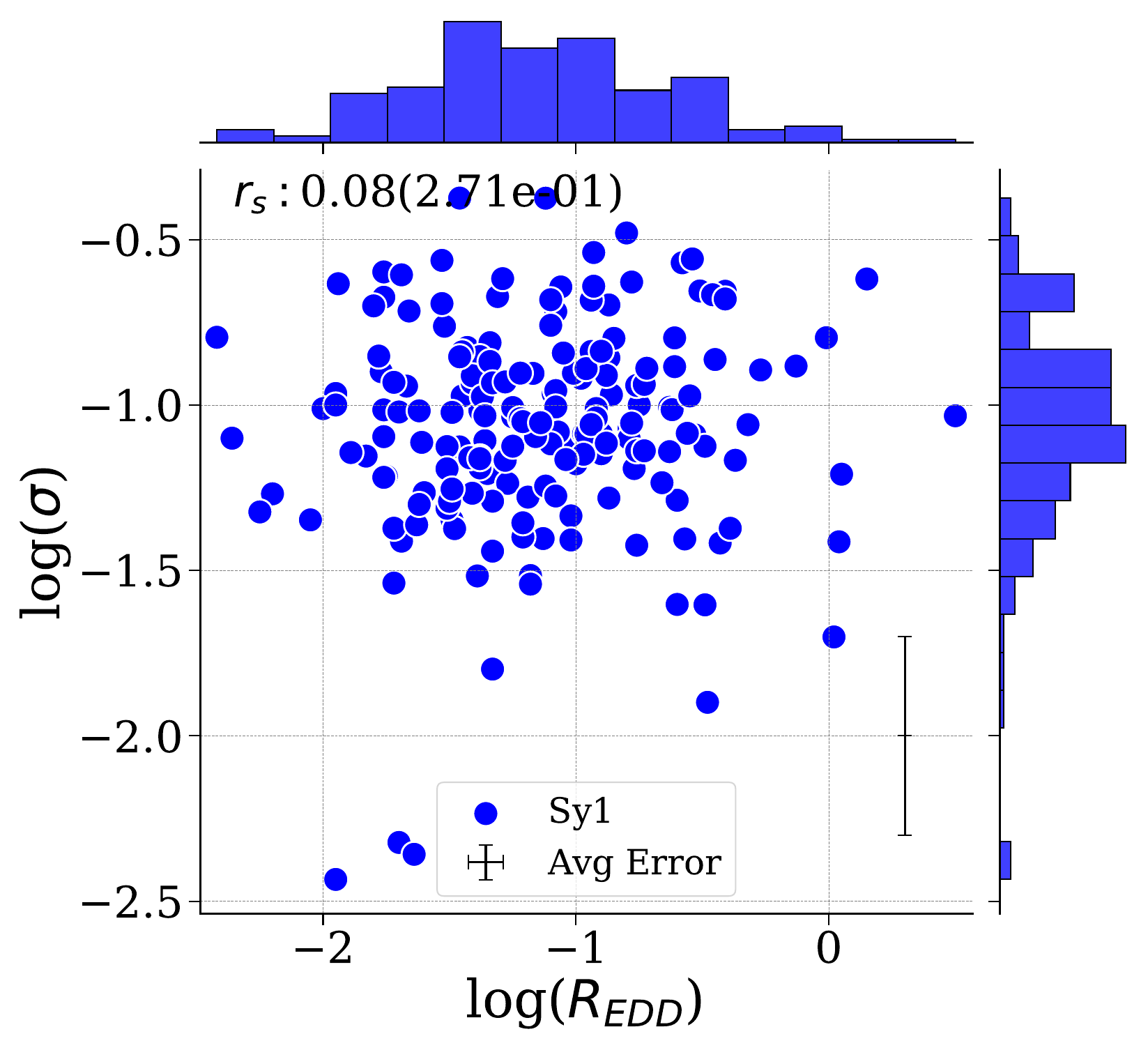}}

\caption{Comparison of the DRW variability amplitude ($\sigma$) with the physical parameters. The left panels show the comparison with $L_{BOL}$; the middle panel shows the comparison with SMBH mass, and the right panel shows the comparison with Eddington ratio $R_{EDD}$. The details are as per Figure \ref{comparison_fvar}.}
\label{sigma_physical_params}
\end{figure*}

\section{Analysis}

\label{section3}

\subsection{Estimation of excess variance}
To quantify the variability in the AGN light curves, we utilise the method described by \citet{vaughan2003} to calculate the fractional variability amplitude,  $F_{\rm var}$. This measure, known as excess variance, provides a measure of the intrinsic variability present in light curves in our sample. We calculate the fractional variability amplitude as:

\begin{equation}
\label{eqn:fvar}
F_{\rm{var}} = \sqrt{ \frac{X^{2} -
\overline{\sigma_{\rm{err}}^{2}}}{\bar{x}^{2}}}.
\end{equation}

Where X is the sample variance calculated as:

\begin{equation}
\centering
X^2=\frac{1}{N-1}\sum_{i=1}^{N}(x_i-\bar{x})^2,
\end{equation}

N is the number of data points in the light curve, and  $\bar{x}$ is the arithmetic mean of the flux value of the light curve. $\overline{\sigma}$ is the mean square error calculated as:

\begin{equation}
\overline{\sigma_{\rm err}^{2}}=\frac{1}{N}\sum_{i=1}^{N}\sigma^{2}_{\rm{err,i}}
\end{equation}

where $\sigma^{2}_{\rm{err,i}}$ is the error in each data point.  $F_{\rm var}$ gives us a measure of how variable the light curve is with respect to the mean.

\subsection{DRW modelling of the light curves }

The DRW model has been frequently employed to model the stochastic light curves of AGN \citep{Kelly2009, MacLeod2010, Zu2011,zu2013}. This model has been successfully applied to various AGN samples, demonstrating correlations between the variability parameters and the physical properties of AGNs. However, some studies suggest that the DRW may not be the underlying process leading to AGN variability and may introduce biases or degeneracies in the parameter estimation. While higher-order Continuous AutoRegressive Moving Average (CARMA) models have been used in a few cases, the DRW model continues to provide a reasonable approximation owing to its simplicity and providing physically meaningful parameters that facilitate comparison with previous works \citep{2022ApJ...936..132Y}. Modelling the light curves as a DRW process yields two key parameters: the damping timescale ($\tau_d$) and the variability amplitude ($\sigma$). The damping timescale is the characteristic time after which the light curve becomes uncorrelated, reflecting the memory of the system (see Figure \ref{fig:drw_acf}). The variability amplitude, on the other hand, is the asymptotic value of the structure function or the root mean square of the flux variations at long timescales. These two parameters can be related to physical parameters of AGNs such as black hole mass, luminosity, Eddington ratio, and rest-frame wavelength.

We use Celerite\footnote{\url{https://github.com/dfm/celerite}}, a Python package for modelling light curves as a DRW process. It provides a fast and efficient way to model time-series data using Gaussian Processes (GPs).  The damping timescales ($\tau_d$) derived from the DRW modelling were converted to the rest frame using the known redshifts of the sources, i.e., $\tau_{d,rest}=\tau_{d,obs}/(1+z)$, z being the redshift of the source. The distribution of the excess variance, variability amplitude and damping timescale is shown in Figure \ref{hist_params}.

To quantitatively assess the impact of both baseline duration and cadence on the accuracy of DRW timescale recovery, we generated artificial DRW light curves with a true damping timescale of $\tau_d$ = 300 days (i.e., log($\tau_d$) = 2.475), we varied the total duration of the light curve (baseline) from 500 days to 3000 days. For each baseline length, we tested four cadence scenarios:

\begin{enumerate}
    \item Regular 1-day sampling
    \item Regular 3-day sampling
    \item Regular 10-day sampling
    \item Realistic seasonal gaps (simulating ZTF-like visibility windows)
\end{enumerate}

For each case, we extracted the median recovered timescale along with the 16th and 84th percentiles. We found that longer baselines significantly improve DRW timescale recovery. For short baselines (e.g., 500–1000 days), the recovered timescales often deviate from the true value and show larger uncertainties. In contrast, light curves with baselines $\geq$ 2000 days generally recover log($\tau$) within $\pm$5\% of the true value across all cadences. Next, we also see the effect of cadence in recovery of the damping timescale. For a fixed baseline, denser sampling (1-day cadence) yields more consistent and accurate recovery of $\tau_d$ while sparse sampling (10-day or seasonal) introduces larger scatter and occasional biases. We examined the real ZTF light curves used in this study (see Figure \ref{fig:sim_drw}). Despite some variation in baseline due to magnitude-dependent detection completeness, the majority of our sources fall within the regime where $\tau_d$ is reliably recovered, as demonstrated by the simulations.

\section{Results}
\label{section4}

\subsection{Optical Variability Characteristics}

Examining the distribution of the variability parameters, we can clearly distinguish between the Type 1 and Type 2 AGN in our sample. First, we analyse the excess variance ($F_{var}$). Type 1 AGN exhibit higher variability, with a median value of $F_{var} = 0.721$, whereas for Type 2 AGN, the median $F_{var}$ is $0.370$. In terms of the DRW parameters, Type 1 AGN have a median log($\sigma$) of $-1.050$, while Type 2 AGN exhibit a median log($\sigma$) value of $-1.873$. The variability parameters obtained through both methods clearly indicate that Type 1 AGN appear significantly more variable than Type 2 AGN. There is a good correlation observed between $\sigma$ and $F_{var}$ (see Figure \ref{comparison_fvar}). Moreover, the median value of the damping timescale ($\tau_d$) for Type 2 AGN is approximately 24.5 days, whereas for Type 1 AGN, the median value is about 295 days. Recent works have associated $\tau_d$ with the thermal timescale of the accretion disk \citep[e.g.,][]{Kelly2009, 2021Sci...373..789B}, suggesting that the shorter timescales observed for Type 2 AGN are likely unphysical. This discrepancy is likely due to the obscuration blocking the flux from the accretion disk continuum, implying that the DRW process may not accurately model Type 2 AGN light curves. We also confirmed this by visually examining the Type 2 AGN light curves, and we found that many of them exhibit minimal variability, possibly rendering the DRW modelling infeasible. Consequently, out of 528 sources, we are left with 303 sources, but it highlights the inability to apply the DRW model to Type 2 AGN.

The ZTF light curves may exhibit spurious variations due to varying contributions from the variations in PSF. To assess the impact of such effects, we selected a control sample of approximately 400 galaxies from the Galaxy and Mass Assembly (GAMA) Survey \citep{2015MNRAS.452.2087L}, ensuring that only passive galaxies without any AGN component were included. For these sources, we computed the excess variance and modelled the light curves using a DRW process, as used for our AGN sample. We found that the median DRW-derived timescale, $\tau_d$, for this passive galaxy sample is 26 days—comparable to that of our Type 2 AGN sample (median $\sim$24 days). In contrast, Type 1 AGNs exhibit significantly longer variability timescales, with a median $\tau_d$ of approximately 295 days. Additionally, the median variability amplitude for the passive galaxy sample is $\log(\sigma) = -1.46$, which is clearly distinct from the Type 1 AGNs ($\log(\sigma) = -1.05$), while for our Type 2 population, median $\log(\sigma$) is -1.87. This distinction in both variability amplitude and timescale between AGNs and passive galaxies suggests that the trends observed in our AGN sample are not dominated by the  artefacts of PSF-induced variability, but rather reflect intrinsic properties of Type 1 AGN variability.

\subsection{Connection between Optical variability and physical parameters}

We obtain the physical parameters for our sample from the BASS-DR2 AGN catalog \citep{koss2022}, which presents the spectroscopic analysis of the entire AGN sample from the 70-month BAT AGN catalog. This sample is comprehensive, encompassing every AGN detected in this X-ray survey. In our study, we find that 353 out of the 528 sources have corresponding data available in the BASS-DR2 catalog. Out of the 303 Type-1 AGN in our sample, we find 227 sources in this catalog. The reduction in sources is due to the use of the 70-month catalog in \citet{koss2022}, whereas our sources are derived from the 105-month catalog. From their work, we select the supermassive black hole (SMBH) mass ($M_{BH}$), the bolometric luminosity ($L_{BOL}$), and the Eddington ratio ($R_{EDD}$) as physical parameters for the Type 1 AGN in our sample.

We  examine the dependence of the variability amplitude ($\sigma$) on the three physical parameters. We found a slight correlation between the SMBH mass and $\sigma$ with a correlation coefficient of 0.24 and a modest positive correlation with $L_{BOL}$, reflected by a coefficient of 0.33. There is almost no dependence of $\sigma$ on $R_{EDD}$, with a correlation coefficient of 0.08  as shown in Figure \ref{sigma_physical_params} (right panel).

Recent studies \citep[e.g.,][]{kozlowski2017, 2021Sci...373..789B, 2023ApJ...951...93C} have demonstrated correlations between AGN variability and physical parameters. To quantify these relationships, we calculate the Spearman rank correlation between the variability and physical parameters. The damping timescale ($\tau_d$) shows a moderate correlation with both SMBH mass and luminosity, with correlation coefficients of 0.35 and 0.40, respectively. However, $\tau_d$ exhibits a weak dependence on the accretion rate, with a correlation coefficient of 0.13. It is important to note that these correlation coefficients are based solely on the Type 1 AGN in our sample, as the timescales obtained for Type 2 AGN appear unphysical due to their heavy obscuration, which likely invalidates the damped random walk (DRW) modelling for these sources.

\begin{figure}

    \includegraphics[width=9cm,height=8cm]{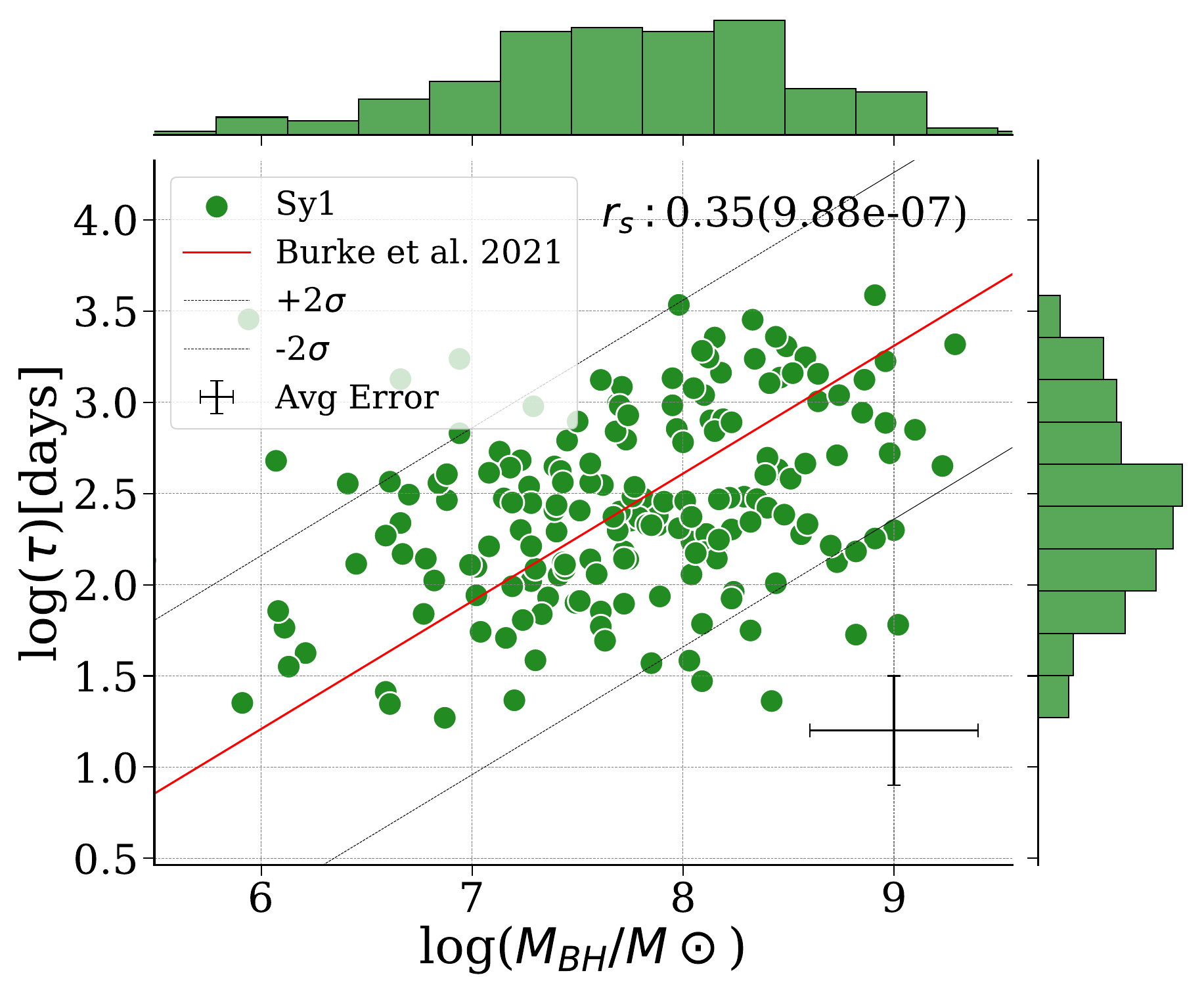}
    \caption{The relation between the SMBH mass and the damping timescale ($\tau_d$) for the Type-1 AGN from our sample. The red line shows the relation obtained in \citep{2021Sci...373..789B}, and the black dashed lines show the 2$\sigma$ deviation from their relation.}
    \label{mbh_tau_plot}
\end{figure}

\citet{2021Sci...373..789B} established a relation connecting the SMBH mass with $\tau_d$. We plot the $M_{BH}$-$\tau_d$ relation for our sources using the relation obtained by them. Our results indicate that, within 2$\sigma$ levels, most sources agree with this relation, as can be seen in Figure \ref{mbh_tau_plot}. However, we acknowledge that the length of the light curves significantly affects the calculation of damping timescales \citep{kozlowski2017}. Therefore, our results, where the light curve lengths are shorter than 10$\tau$, should be considered as upper limits which will improve as longer light curves become available.

\subsection{Connection between optical variability and  X-ray parameters}

\begin{figure*}
\hspace{-1cm}
    \subfigure{\includegraphics[width=9cm,height=8cm]{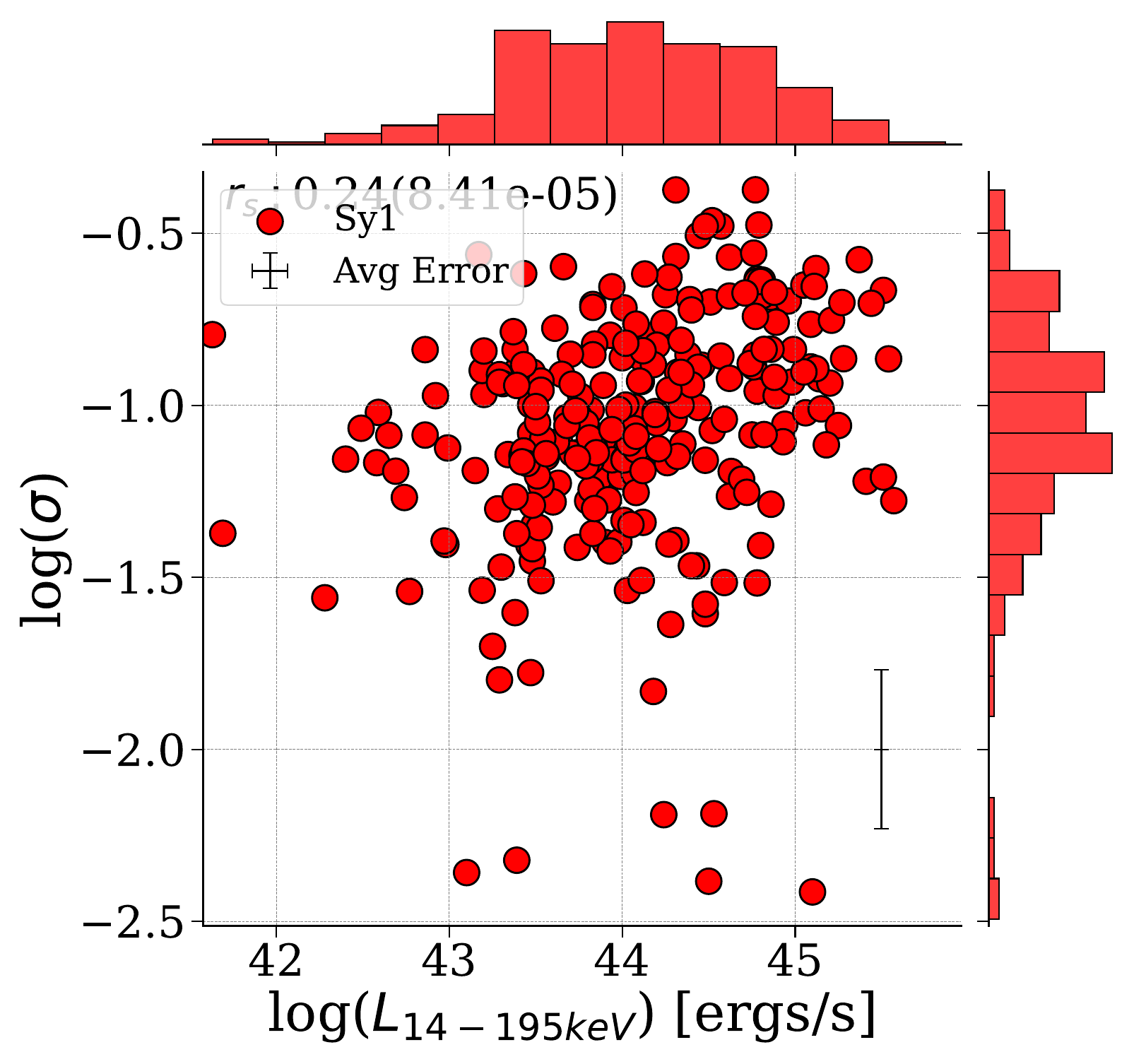}}
    \subfigure{\includegraphics[width=9cm,height=8cm]{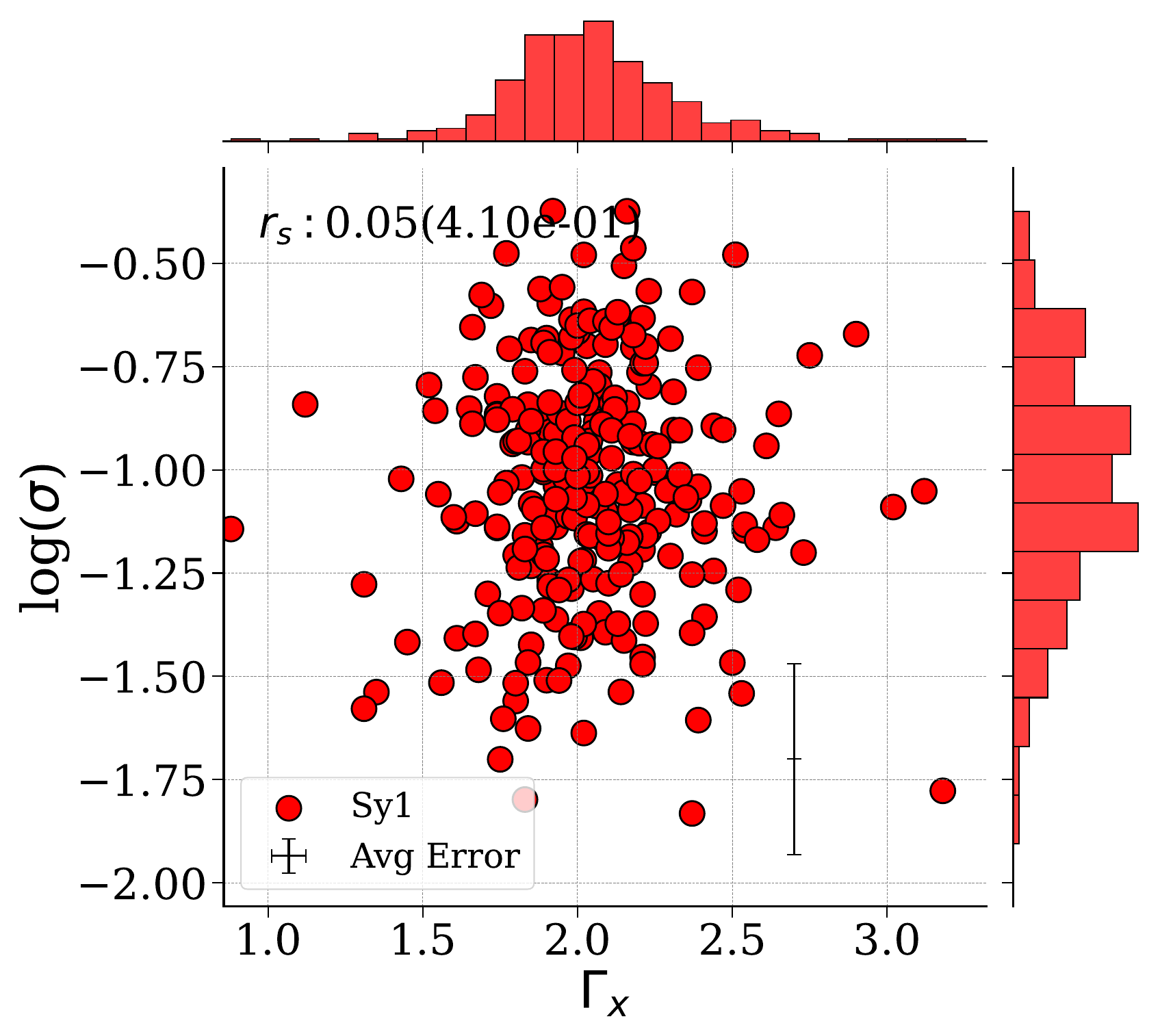}}
    \caption{The relation between the DRW variability amplitude ($\sigma$) and the 14-195 KeV X-ray flux for the Type-1 AGN from our sample in the top panel and the X-ray photon indices ($\Gamma_x$) in the bottom panel. The details are as per Figure \ref{comparison_fvar}.}
  \label{xray_parameters}
\end{figure*}

We obtain the X-ray parameters from the BAT 105-month catalogue\footnote{\url{https://swift.gsfc.nasa.gov/results/bs105mon/}}, utilising the photon indices ($\Gamma_X$) and X-ray flux in the 14–195 \text{keV} range. In systems with accreting compact objects, such as black holes or neutron stars, the photon index varies due to changes in accretion rates, magnetic fields, and the geometry of the accretion flow \citep{1993ARA&A..31..717M, 2003ApJ...593...96M}. Optical variability often mirrors these X-ray changes; as matter spirals toward the compact object, it heats up, emitting X-rays. The optical emission can then respond to these variations in the accretion process, as observed in multi-wavelength studies of AGN \citep{2003ApJ...584L..53U, 2006Natur.444..730M}. Furthermore, the X-ray luminosity quantifies the total energy output in X-rays, which depends on the accretion rate and the efficiency of energy conversion \citep{1994ApJS...95....1E, 2009MNRAS.399.1597S}. Variations in X-ray luminosity often coincide with changes in the accretion state, impacting the surrounding environment and affecting optical emission. Conversely, optical variations can alter the accretion rate, thereby impacting X-ray luminosity. Investigating these feedback mechanisms sheds light on the dynamic processes governing these systems \citep{2012ARA&A..50..455F}.

Our analysis reveals a negligible correlation between $\sigma$ and $\Gamma_X$, with a correlation coefficient of 0.046, while a correlation coefficient of 0.24 between the X-ray luminosity and $\sigma$ was observed, indicating slight dependence as shown in Figure \ref{xray_parameters}. Additionally, we do not find any dependence of these parameters on the damping timescale $\tau_d$, indicating that the X-ray parameters may not significantly drive the observed long-term optical variability of AGN. 

\begin{figure}

\includegraphics[width=9cm,height=8cm]{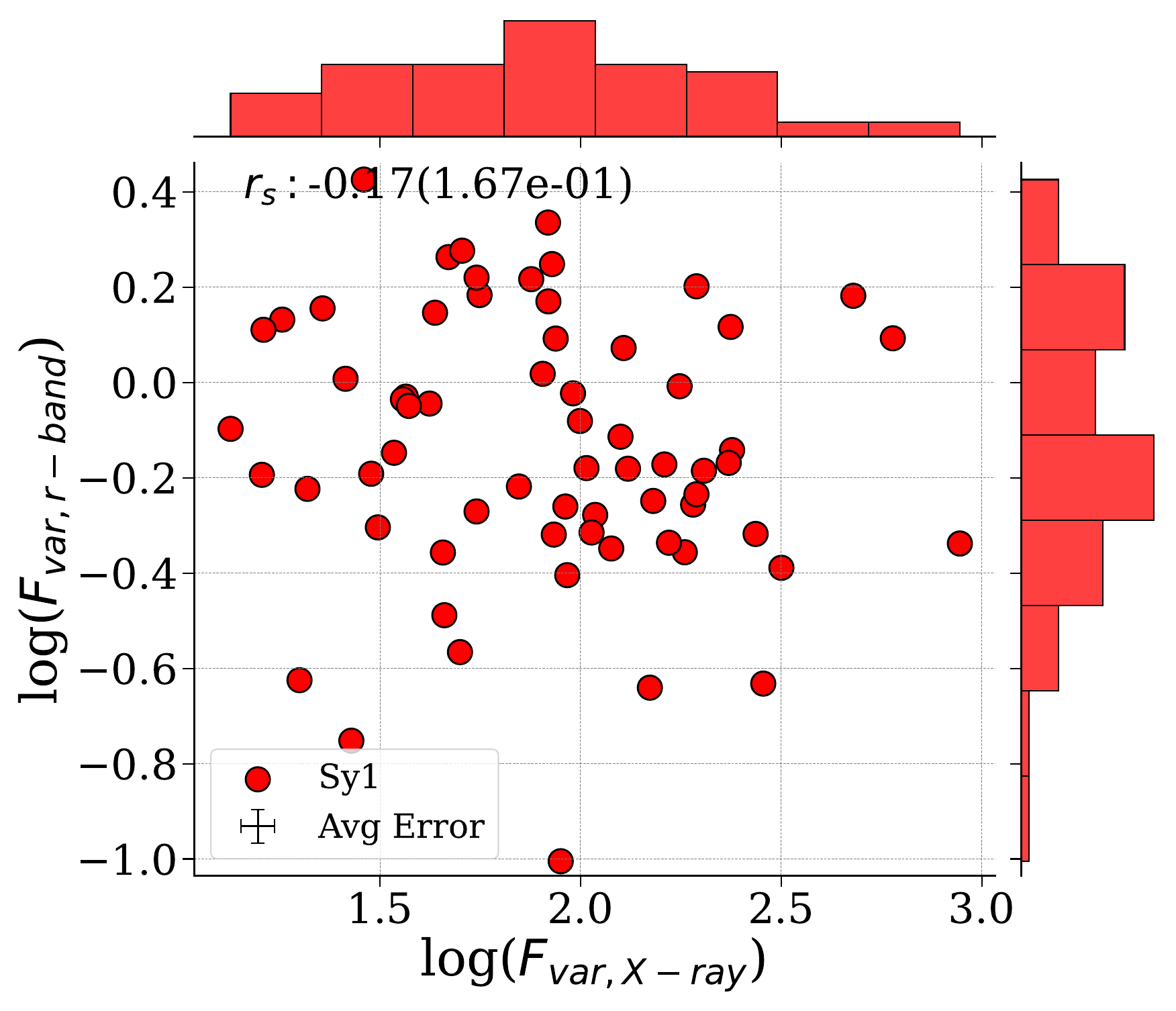}

\caption{Comparison of the variability parameters obtained in the r band (x-axis) and in X-ray (y-axis). The details are as per Figure \ref{comparison_fvar}.}
\label{variability_opt_xray}
\end{figure}

 \citet{2020ApJ...896..122L} studied the BAT AGN 105-month sample in order to search for Supermassive Black Hole Binaries. They used the X-ray light curves and calculated the excess variance in these light curves. We use their values to get the long-term X-ray variance and compare it with the optical variability characteristics calculated in the previous section. For this, we cross-matched our 303 sources with their sample and identified 82 Sy1 sources common in both studies. Interestingly, we find a weak anti-correlation between the optical and X-ray variability trends, with an anti-correlation coefficient of $-0.17$ (see Figure \ref{variability_opt_xray}). It is worth noting that the light curves used for calculating X-ray variability span approximately 5-6 years, whereas our variability trends are also based on about 5-year-long optical light curves, indicating similar timescales being probed, although non-simultaneous. As earlier studies have shown,  the timescale of X-ray variability is much shorter than optical timescales and short-term variations from the inner regions could affect long-term variations in the outer regions \citep{2010MNRAS.403..605B}. However, if the disk emission is indeed being reprocessed, we would expect to see a correlated variability trend between the two wavelengths, and sources with higher X-ray variability should be highly variable in optical bands too \citep{2014ApJ...788...48S, 2015ApJ...806..129E} which we did not observe for the 82 Sy1 galaxies studied here.

\subsection{ Connection between optical variability and radio measurements}

We search for the sources in the radio catalogues, specifically the Faint Images of the Radio Sky at Twenty-Centimeters (FIRST) catalogue, which consists of 1.4 GHz radio observations \citep{1995ApJ...450..559B}. From this catalogue, we obtain the peak flux as well as the integrated flux for each source. Out of 303 Type 1 AGN, we identify 87 sources with available radio data, providing a unique opportunity to study the radio, optical, and X-ray flux variability simultaneously. 

\begin{figure*}
\hspace{-1cm}
    \subfigure{\includegraphics[width=9cm,height=8cm]{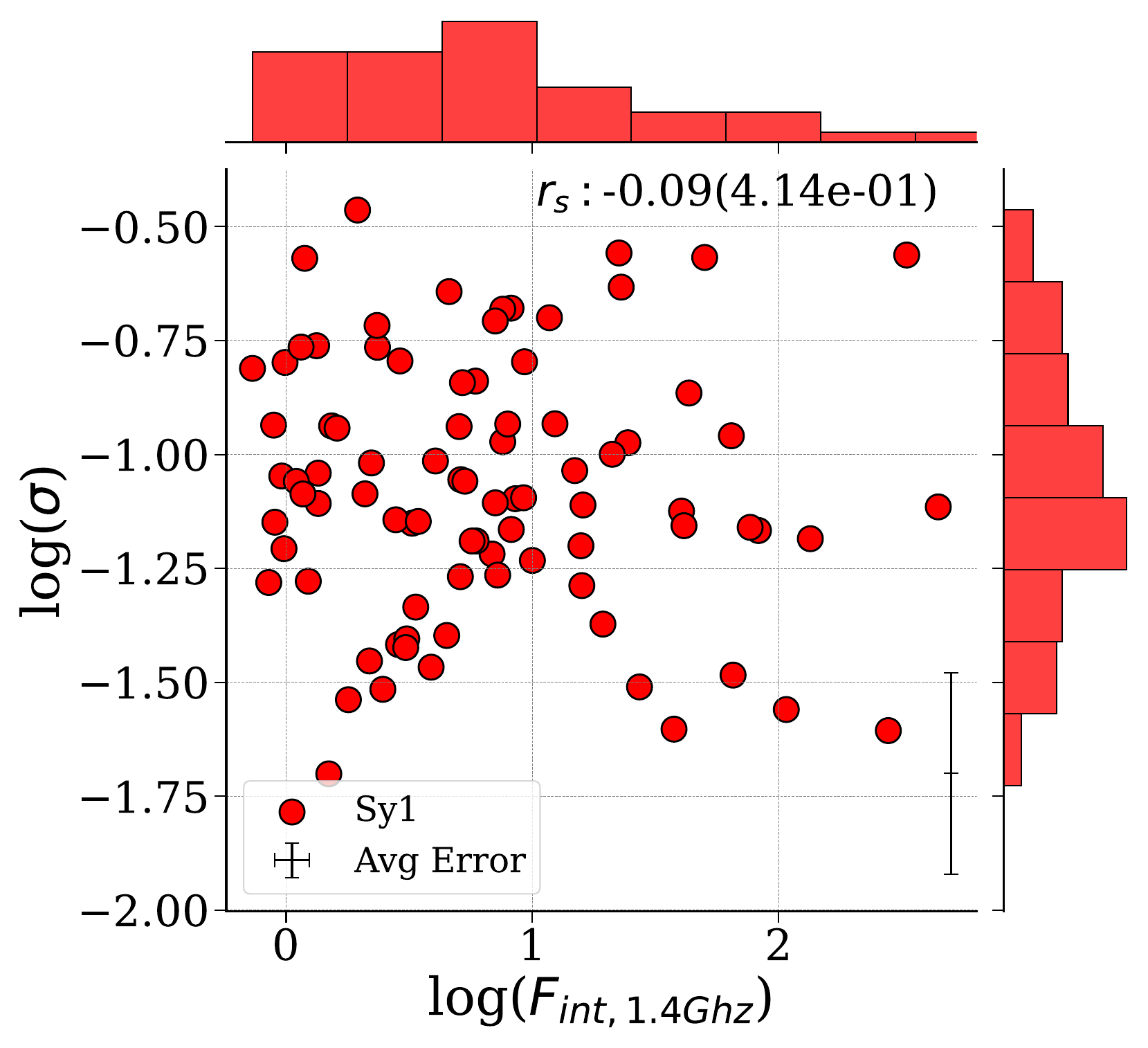}}
    \subfigure{\includegraphics[width=9cm,height=8cm]{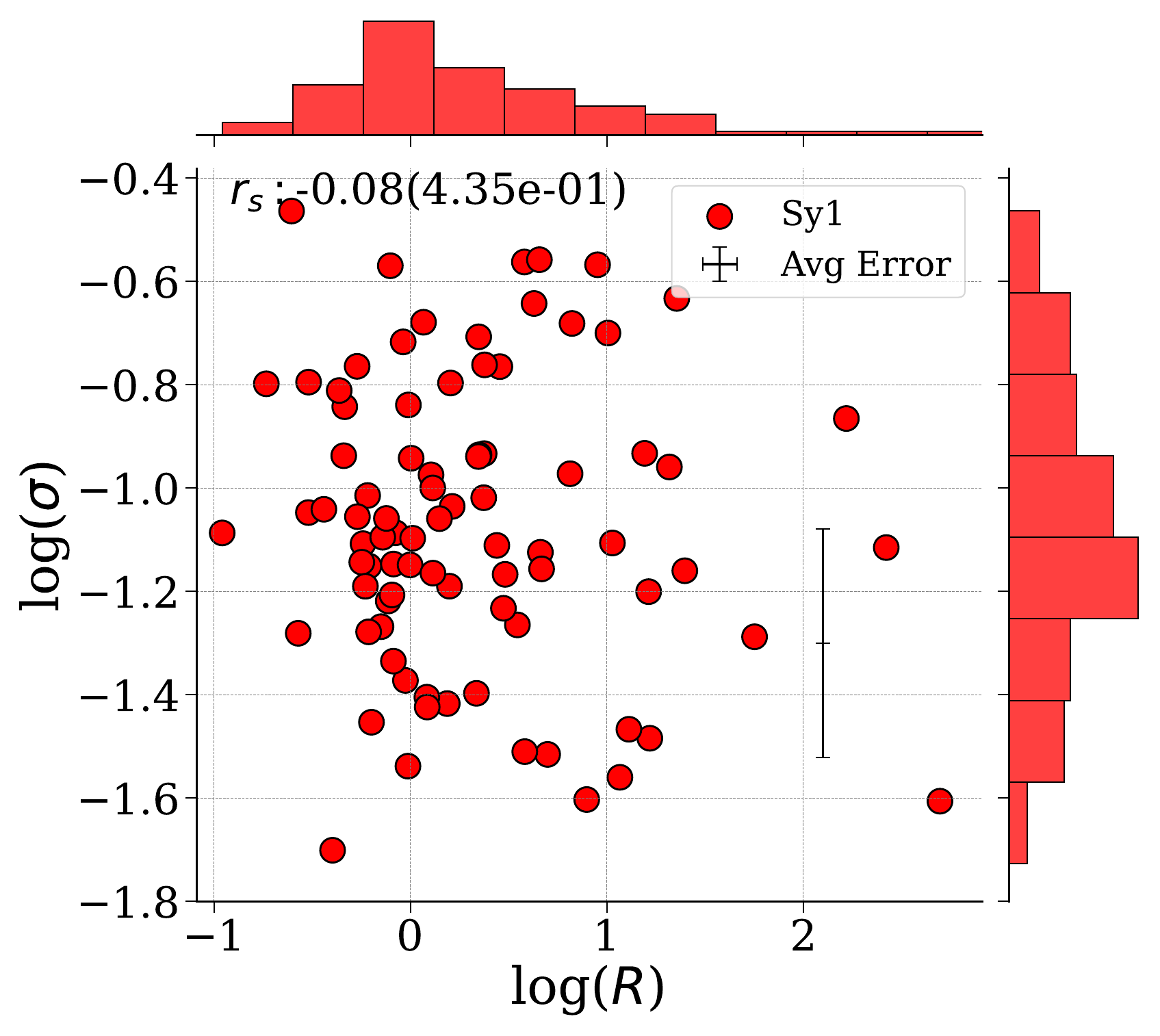}}
    \caption{The relation between the DRW variability amplitude ($\sigma$) and the 1.4 Ghz integrated radio flux for the Type-1 AGN from our sample in the top panel and the Radio Loudness in the bottom panel. The details are as per Figure \ref{comparison_fvar}.}
  \label{radio_parameters}
\end{figure*}

Radio emission in AGN often results from the jet activity, suggesting that radio-loud sources, which are typically associated with strong jet emissions, might be expected to exhibit higher variability \citep{1983ApJ...266...18U, 1994A&A...289..673T,1995AJ....110..529C}.  Our analysis reveals almost no correlation between the peak flux at 1.4 GHz and the variability amplitude $\sigma$, with a correlation coefficient of $-0.09$. Additionally, we calculate the radio loudness parameter (RL) for these sources, defined as the radio to optical flux ratio, to examine any optical variability dependency on radio loudness \citep{1994AJ....108.1163K}. We find almost no correlation between RL and $\sigma$ with an anti-correlation coefficient of $-0.08$ (Figure \ref{radio_parameters}). Our findings suggest that the relationship between radio loudness and optical variability may be complex and might be influenced by additional factors such as the orientation of the jet, the structure of the accretion disk, and the intervening medium.

Also, since the radio and optical emissions probed in this work are non-simultaneous, it would be interesting to test whether there is any correlation between the two wavelengths in the case of simultaneous observations. Future studies with larger samples and more detailed radio observations will be helpful to understand further the connection between radio emission and optical variability in AGN. The integration of data from new surveys, such as the Very Large Array Sky Survey (VLASS) and the upcoming Square Kilometre Array (SKA), will be particularly valuable in advancing our understanding of these mechanisms.

\begin{table*}
\centering
    \caption{\bf Correlation Table with Spearmann Rank Coefficients (null values denoted inside the bracket)}
    \begin{tabular}{lcccc}
        \toprule
      Parameter & Excess Variance ($F_{var}$) & Sigma ($\sigma$) & Tau ($\tau$)\\
        \hline
        $M_{\text{BH}}$         & 0.072 (0.27) & 0.242 (10$^{-3}$) & 0.352 (10$^{-07}$) \\
        $L_{\text{bol}}$        & 0.141 (0.03) & 0.33 (10$^{-06}$) & 0.397 (10$^{-10}$) \\
        $R_{EDD}$                & 0.093 (0.16) & 0.08 (0.02) & 0.131 (0.05) \\
        $R_{\text{int}}$       & $-$0.094 (0.34) & $-$0.102 (0.30) & $-$0.032 (0.74) \\
        $R_{\text{peak}}$        & $-$0.141 (0.15) & $-$ 0.09 (0.41) & $-$0.069 (0.49) \\
        $R_{\text{loudness}}$   & $-$0.201 (0.04) & $-$0.08 (0.43) & 0.096 (0.33) \\
        $\Gamma_X$                & 0.017 (0.77) & 0.046 (0.42) & 0.056 (0.33) \\
        $L_{\text{X}}$          & 0.12 (0.02) & 0.24 (0.03) & 0.065 (0.04) \\
        $\sigma_{X-ray}$       & $-$0.175 (0.10) & $-$0.15 (0.22) & $-$0.279 (10$^{-3}$) \\
        \hline
    \end{tabular}

    \label{correlation_table}
\end{table*}

\section{Discussion}
\label{section5}

Through this work, we probe the optical variability signatures of AGN in the 105-month Swift-BAT catalog, employing optical light curves spanning approximately five years from the ZTF survey. Variability trends were quantified via excess variance metrics, and light curves were modelled employing a DRW formalism, from which we extracted the DRW timescale, $\tau_{\mathrm{d}}$, and variability amplitudes, $\sigma_{\mathrm{DRW}}$. Our analysis presents variability distinctions between Type 1 and Type 2 AGN. In Type 2 AGN, optical variability is markedly suppressed due to obscuration effects, rendering them less variable in appearance compared to their unobscured Type 1 counterparts. This observation lends empirical support to the unified AGN model, implying that observed discrepancies are predominantly attributable to orientation relative to the obscuring torus. This will be useful in distinguishing between Type 1 and Type 2 AGN, especially in the era of Rubin/LSST, where similarly sampled light curves will be available for a large number of sources, many of them without spectroscopic information.

\citet{2023MNRAS.518.1531L} presented a systematic study of optical variability in spectroscopically classified Type 1 and Type 2 AGNs using ZTF data.Their analysis demonstrated that DRW-derived parameters, such as longer damping timescales and higher asymptotic structure function amplitudes ($\mathrm{SF}_{\infty, g} > 0.07$ mag), correlate with the presence of broad H$\alpha$ emission, helping to uncover misclassified or transitional sources. Our findings, particularly the distinct $\tau_d$ and variability amplitude distributions between Type 1 and Type 2 AGNs, are consistent with their results and further reinforce the diagnostic power of time-domain analysis in identifying AGN sub-classes and identifying outliers within the unified scheme.

While we adopt the DRW model as our primary framework for characterising AGN optical variability, we acknowledge that it does not universally constitute the best statistical description of every light curve in our sample. Nonetheless, our modelling results suggest that the DRW process is broadly capable of capturing the dominant variability structure in the majority of AGN light curves. In this context, the interpretability and compactness of the DRW parametrisation, yielding characteristic values of variability amplitude ($\sigma$) and timescale ($\tau_d$), remain valuable for comparing across populations and linking variability to AGN physical properties.

We acknowledge, however, that more flexible statistical frameworks, may provide improved descriptions in specific cases, particularly for light curves exhibiting complex memory structure or low-level nonstationarity \citep[see][]{Kozz2016,2022ApJ...936..132Y,2025A&A...693A.319K}. Here, our aim is to provide a tractable yet physically interpretable characterisation of variability across a large and diverse AGN sample.

Table \ref{correlation_table} presents correlation coefficients for all analysed parameters. We find a positive correlation between the DRW timescale and both SMBH mass and AGN luminosity. This relationship intimates that AGNs harbouring more massive black holes and exhibiting higher luminosity—presumably hosting more expansive accretion disks, demonstrate protracted characteristic variability timescales. Such a correlation is consistent with theoretical models where the depth of the gravitational potential well and accretion disk dimensions critically influence variability timescales. and is consistent with the recently obtained results 

However, our correlations between the variability and physical parameters are weaker than those reported in recent works \citep[see][etc.]{2021Sci...373..789B, 2024A&A...684A.133A}. A plausible explanation for this likely discrepancy with previous findings may lie in the nature of our sample- it is drawn from a hard X-ray selected AGN population, which might exhibit different accretion and obscuration properties compared to optically selected samples. Such selection may introduce biases in black hole mass and Eddington ratio distributions, thereby diluting their observed expected trends in our sample.

We extended our study to investigate the potential correlation of X-ray emission with long-term optical variability using parameters such as spectral indices, luminosity, and variability patterns. \citet{2018ApJ...868...58K} identified correlations between UV/optical variations and X-ray loudness defined as X-ray luminosity to bolometric luminosity in quasars. However, we find no statistically significant correlation between optical variability and either X-ray luminosity or hard X-ray photon indices in our AGN sample, which does not agree with their results, indicating the relation between X-ray luminosity and AGN variability remains to be clearly established.

Our analysis reveals  anti-correlation between long-term optical and X-ray variability amplitudes (Figure \ref{variability_opt_xray}). This indicates a difference between emission mechanisms originating from distinct accretion disk regions, the X-ray-emitting hot inner disk and the optically-emitting cooler outer regions. The absence of any significant correlation demonstrates that optical variability mechanisms operate independently of X-ray emission processes, consistent with recent studies that report only modest UV/optical-X-ray correlations in AGNs. This evidence challenges conventional disk reprocessing models.

While statistically modest, the detected anti-correlation between  X-ray and optical variability amplitudes supports two distinct physical scenarios: (1) independent variability mechanisms dominate different disk regions, or (2) intrinsic thermal disk fluctuations, rather than high-energy photon reprocessing, drive the observed variability. \citet{2008MNRAS.389.1479A} explored the correlated variability of optical and X-ray emissions in AGNs. Their findings suggest that the variability is driven both by the reprocessing as well as the accretion rate fluctuations. \citet{2009MNRAS.394..427B} investigated the correlation between X-ray and optical variability in AGNs. They found that although the X-ray and optical light curves appear correlated, the optical variations possibly arise from a different mechanism other than reprocessing. \citet{2024MNRAS.528.5972P} studied the long-term X-ray and optical variability of SDSS quasars using data from the SRG/eROSITA all-sky survey. They concluded that the X-ray variability anti correlates with the SMBH mass and lumiunosity.   Recently, \citet{2024MNRAS.530.4850H} concluded that the optical/UV variability in the disk is intrinsically driven and not by reprocessing of the X-ray photons. \citet{2024ApJ...975..160R} studying the same sample as ours found out that the disk fluctuations indeed cause the AGN variability seen in the optical light curves. In this situation, if no significant correlation between the optical variability and X-ray parameters or even the X-ray variability is found, it supports the argument that the long optical variability in AGN accretion disks is independent of the X-ray emission from AGN.

We also study the effect of radio emission in causing optical variability by correlating the radio fluxes and loudness (utilizing the FIRST 1.4 GHz survey) with optical variability. Short-term optical variability has been seen in sources with jets, and emission dominated by the jets may cause this phenomenon. For instance, Blazars, with jets pointed towards us and the NLSy1 galaxies with confirmed jets, are highly variable on short timescales if they are radio loud \citep{2018BSRSL..87..281G,2022MNRAS.514.5607O}. We observe an anti-correlation between optical variability, radio flux, and radio loudness. Our finding of an anti-correlation between optical variability and radio flux suggests that the processes governing optical emissions, likely related to the accretion disk, are distinct from those driving radio emissions in the jet. This finding aligns with the notion that radio-loud AGNs, which exhibit stronger and more variable emissions, may have different accretion and jet-launching mechanisms compared to their radio-quiet counterparts.  

Future investigations with larger samples and more extended monitoring periods will be helpful to confirm these correlations and refine our understanding of AGN variability. Expanding the sample size will enhance the statistical robustness of the observed trends, while more extended monitoring periods will provide better constraints on the variability timescales and amplitudes.

\section{Conclusions}

\label{section6}

We studied the optical variability of AGN from the 105-month {\it Swift/} BAT catalogue. We characterised the light curves using excess variance and Damped Random Walk (DRW) modelling and investigated correlations between variability parameters and physical parameters, namely the supermassive black hole (SMBH) mass, luminosity, and Eddington ratio. We also analysed the relationship between optical variability and radio and X-ray parameters. We summarise the primary findings of our research as follows:

\begin{enumerate}
    \item Our observations reveal distinct variability trends between Type 1 and Type 2 AGN, supporting the unification model of AGN. As predicted, the obscuration in Type 2 AGN significantly dampens their optical variability, making them appear significantly less variable compared to their unobscured counterparts.

    \item Within our Type 1 AGN subsample, the observed dependence of the damping timescale on both supermassive black hole (SMBH) mass and AGN luminosity aligns with the predictions of accretion disk theory. This suggests that the damping timescale reflects the disk's thermal timescale, where the characteristic timescale for radiating away thermal energy is influenced by both the central black hole's mass and the AGN's overall luminosity.
    
    \item Our study indicates a weak anti-correlation between variability and radio flux as well as radio loudness, suggesting that jet interaction may not influence or correlate with the long-term optical variability trends.
    
    \item We observe a weak correlation between variability and X-ray parameters, including X-ray spectral photon indices and luminosity. This implies that reprocessing of the X-ray photons may not be driving long-term AGN variability but rather the inhomogeneous disk itself.
     
    \item An anti-correlation between optical and X-ray variability amplitudes, even though non-simultaneous, implies a little connection between the long-term variability mechanisms in the hot inner regions of the accretion disk (traced by X-rays) and the cooler outer regions (traced by optical emissions). This highlights the need for comprehensive models that can account for these differences across various wavelengths.
\end{enumerate}

With the availability of large datasets in multiple wavelengths, future studies will explore these aspects of AGN variability in greater detail, potentially uncovering relationships between physical parameters and contributing to the development of robust models that integrate multi-wavelength observations.

\section*{Acknowledgements}

Based on observations obtained with the Samuel Oschin 48-inch Telescope at the Palomar Observatory as part of the Zwicky Transient Facility project. ZTF
is supported by the National Science Foundation under Grant No.
AST-1440341 and collaboration including Caltech, IPAC, the Weizmann Institute for Science, The Oskar Klein Center at Stockholm University, the University of Maryland, the University of Washington,
Deutsches Elektronen-Synchrotron and Humboldt University, Los
Alamos National Laboratories, the TANGO Consortium of Taiwan,
the University of Wisconsin at Milwaukee, and Lawrence Berkeley
National Laboratories. Operations are conducted by COO, IPAC, and
UW.
\\

{\it Facilities: Swift-BAT, ZTF}

\appendix \label{app:a}
\restartappendixnumbering
\section{The simulations with the Damped Random Walk}

\begin{figure}
    \centering
    \includegraphics[width=18cm, height=10cm]{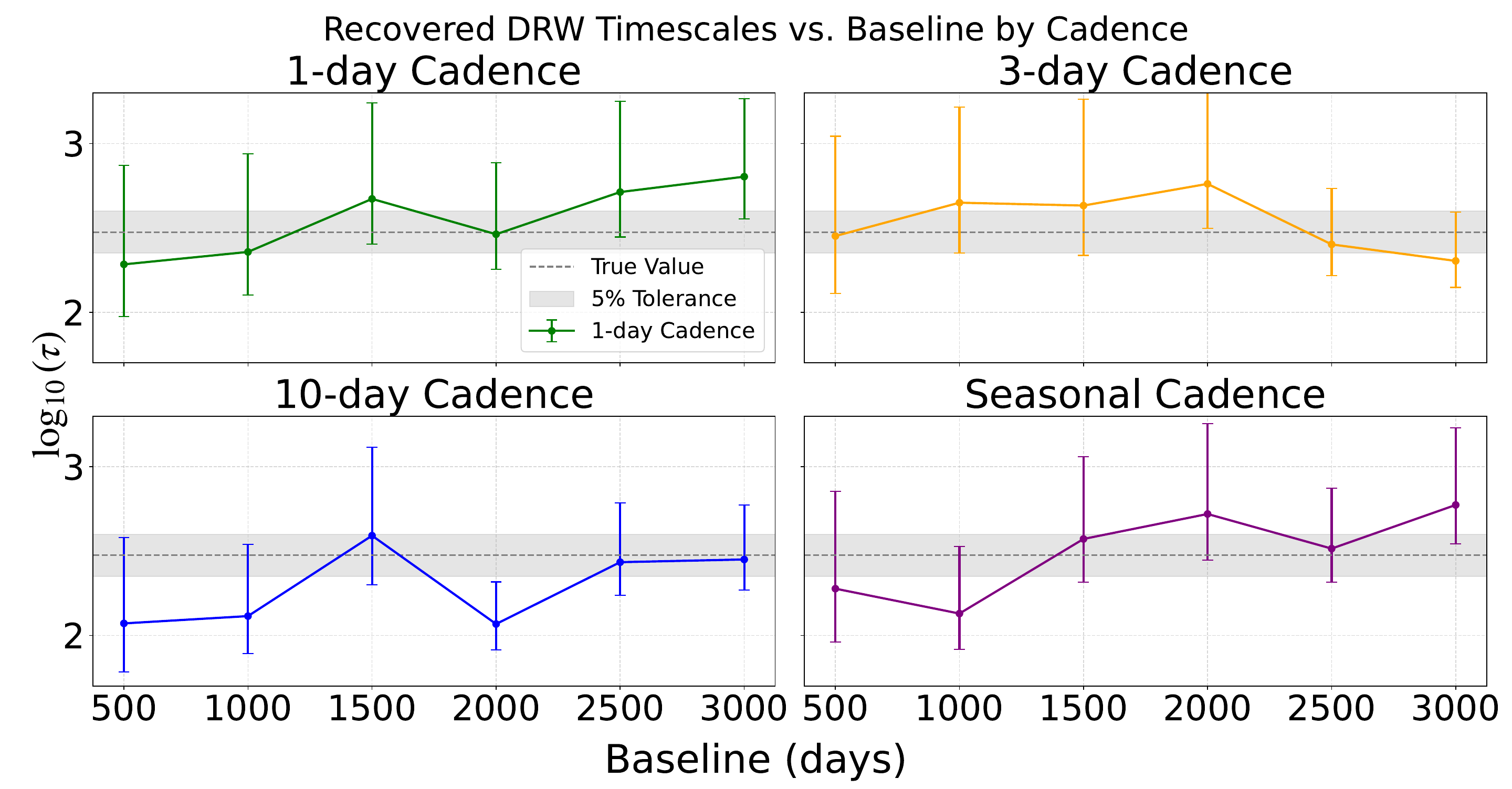}
    \caption{We performed simulation using the DRW process. Four different cadences alongwith various sampling duration are displayed here. For an input $\tau_d$ of 300 days, the length of light curves affect the recovery. The length of ZTF light curves $\sim$ 2000 days used here  can recover $\tau_d$ within the expect error range across the various cadences. This simulation with various cadences including seasonal gaps matching ZTF shows modest scatter but no systematic bias in the recovered parameters.}
    \label{fig:sim_drw}
\end{figure}

\section{Testing the validity of DRW model}
\begin{figure}
    \centering
    \includegraphics[width=18cm, height=11cm]{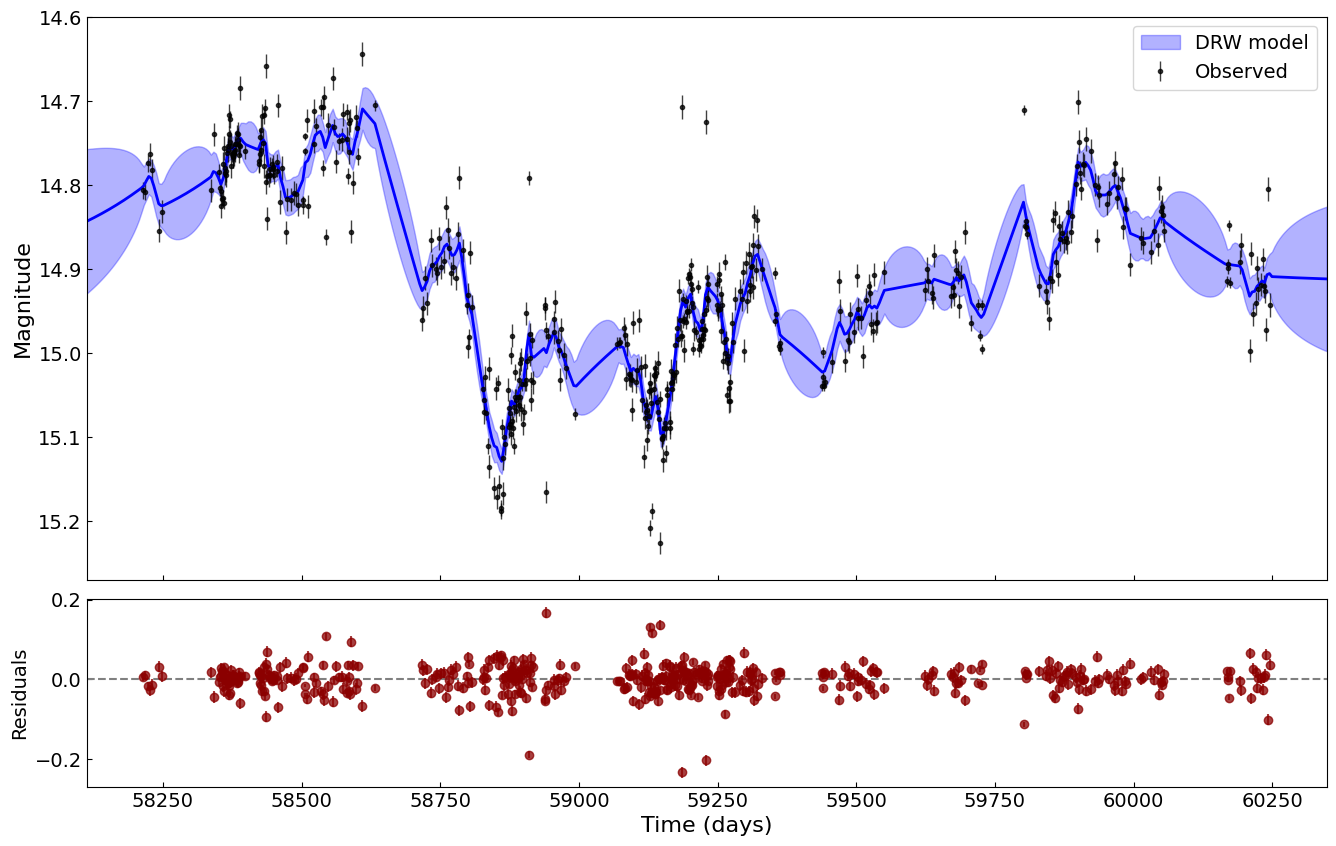}
    \includegraphics[width=18cm, height=7cm]{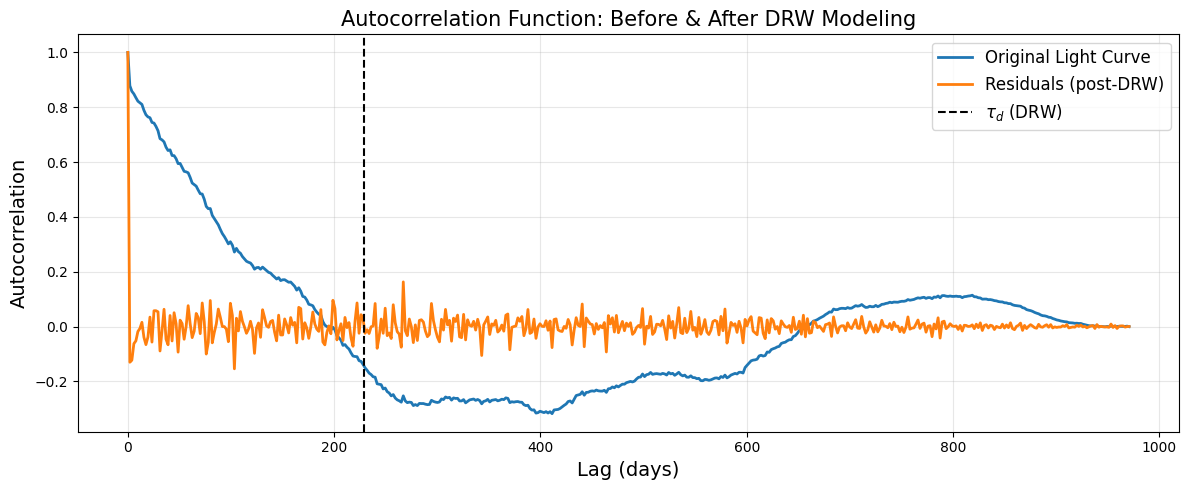}
    \caption{Demonstration of the DRW modeling of the light curve for one of the sources in our sample (R.A.:104.91$^\circ$ and Dec.:54.19$^\circ$ (top panel). The autocorrelation function shown for the light curve before the DRW modeling (blue color) and after the DRW modeling (orange) is shown in the lower panel. }
    \label{fig:drw_acf}
\end{figure}

\bibliographystyle{aasjournal}
\bibliography{main}

\label{lastpage}
\end{document}